\documentclass[english,aps,prx,superscriptaddress,twocolumn,address,showacknowledgments,longbibliography]{revtex4-1}
\usepackage[T1]{fontenc}
\usepackage[latin9]{inputenc}
\usepackage{amsmath}
\usepackage{graphicx}
\usepackage{esint}
\usepackage[normalem]{ulem}

\usepackage{color,colordvi}
\definecolor{blue}{rgb}{0,0,1}
\definecolor{grey}{rgb}{0.6,0.6,0.6}

\makeatletter

\@ifundefined{textcolor}{}
{%
 \definecolor{BLACK}{gray}{0}
 \definecolor{WHITE}{gray}{1}
 \definecolor{RED}{rgb}{1,0,0}
 \definecolor{GREEN}{rgb}{0,1,0}
 \definecolor{BLUE}{rgb}{0,0,1}
 \definecolor{CYAN}{cmyk}{1,0,0,0}
 \definecolor{MAGENTA}{cmyk}{0,1,0,0}
 \definecolor{YELLOW}{cmyk}{0,0,1,0}
 \definecolor{darkorange}{rgb}{1,0.5,0}

}

\newcommand{\pd}{\phantom{\dagger}}

\newcommand{\bk}{\mathbf{k}}
\newcommand{\bj}{\mathbf{j}}

\newcommand{\von}{\nu_{\rm on}}
\newcommand{\voff}{\nu_{\rm off}}

\newcommand{\halpha}{\hat{\alpha}}

\usepackage{bbm}

\makeatother

\usepackage{babel}
\begin{document}


\title{Topological phase transitions and chiral inelastic transport\\ induced by the squeezing of light}

\author{Vittorio Peano}
\affiliation{Institute for Theoretical Physics, University of Erlangen-N\"{u}rnberg,  Staudtstr. 7, 91058 Erlangen, Germany}

\author{Martin Houde}
\affiliation{Department of Physics, McGill University, 3600 rue University, Montreal, Quebec, H3A 2T8, Canada}

\author{Christian Brendel }
\affiliation{Institute for Theoretical Physics, University of Erlangen-N\"{u}rnberg,  Staudtstr. 7, 91058 Erlangen, Germany}

\author{Florian Marquardt}
\affiliation{Institute for Theoretical Physics, University of Erlangen-N\"{u}rnberg,  Staudtstr. 7, 91058 Erlangen, Germany}
\affiliation{Max Planck Institute for the Science of Light, G\"{u}nther-Scharowsky-Stra{\ss}e 1/Bau 24, 91058 Erlangen, Germany}

\author{Aashish A. Clerk}
\affiliation{Department of Physics, McGill University, 3600 rue University, Montreal, Quebec, H3A 2T8, Canada}

\date{\today}

\begin{abstract}
We show how  the squeezing of light can lead to the formation of  topological states.   Such states are characterized by non-trivial
Chern numbers, and exhibit protected edge modes which give rise to chiral elastic and inelastic photon transport.  These  topological bosonic states  are not equivalent to their fermionic (topological superconductor) counterparts and  cannot be mapped by a local transformation onto  topological states found in particle-conserving models.  They thus represent a new  type of topological system.  We study this physics in detail in the case of a Kagome lattice model, and discuss possible realizations using nonlinear photonic crystals or superconducting circuits.
\end{abstract}

\pacs{fill in}
\maketitle

\section{Introduction}  
There has been enormous interest in
trying to replicate the physics of topological electronic phases in a variety of 
bosonic systems, including  cold atoms \cite{Goldman2014}, photonic systems \cite{Lu2014} and more recently phononic systems \cite{Prodan2009,Kane2013,Peano2015,Yang2015,Susstrunk2015,Paulose2015,Nash2015}.  
Photonic analogues include quantum-Hall like states
induced through the introduction of synthetic gauge fields \cite{Raghu2008a,Raghu2008b,Koch2010,Umucallar2011,Fang2012,Petrescu2012,Schmidt2015},
 phases which are analogous to time-reversal invariant topological
insulators \cite{Hafezi2011,Khanikaevphotonic2012,Hafezi2013}, Floquet topological insulators \cite{Kitagawa2012,Rechtsman2013} and even Majorana-like modes \cite{Bardyn2012}.
Experimental studies of such phases have made significant progress
\cite{Wang2009,Rechtsman2013,Rechtsman2013b,Hafezi2013,Mittal2014,2014_Lipson_NonreciprocalPhaseShift}. 
In addition to being of fundamental interest, these topological
photonic and phononic phases could have practical utility, as they
provide disorder-protected edge modes that could be used for chiral
light and sound propagation.

Despite this intense activity, most works on topological bosonic states amount in the end
to replicating a well-known fermionic single-particle Hamiltonian
with bosons; as the topological properties are a function of the resulting
single-particle eigenstates, particle statistics play no crucial role,
except perhaps in the methods used for probing the system. As we now discuss, this simple
correspondence will fail if the particle number is not conserved.  

Consider the most general quadratic Hamiltonian describing bosons on a lattice which respects the discrete translational invariance of the lattice, but which does not conserve particle number:
\begin{equation}
	\hat{H} = \sum_{\bk,n} \varepsilon_n[\bk] b^\dagger_{\bk,n} b^{\pd}_{\bk,n} +
		\sum_{\bk,n,n'} \left( \lambda_{nn'}[\bk] b^\dagger_{\bk,n} b^\dagger_{-\bk,n'} + h.c. \right)
		\label{eq:GenericH}
\end{equation}
The first term describes a non-interacting band-structure, where $\bk$ runs over momenta in the first Brillouin zone, and $n$ labels the bands.  The remaining terms correspond to parametric driving or two-mode squeezing terms.  
As we discuss below, they can be controllably realized in a number of different settings, with the $\hat{b}_{\bk,m}$ operators describing "real" particles 
(i.e.~photons in a cavity lattice), and not quasiparticles defined above some effective condensate.
While superficially similar to pairing terms in a superconductor, these two-mode squeezing terms have a profoundly different effect in a bosonic system, as there is no limit to the occupancy of a particular single-particle state.  They can give rise to highly entangled ground states, and even to instabilities.  
\begin{figure}
	\includegraphics[width=0.9\columnwidth]{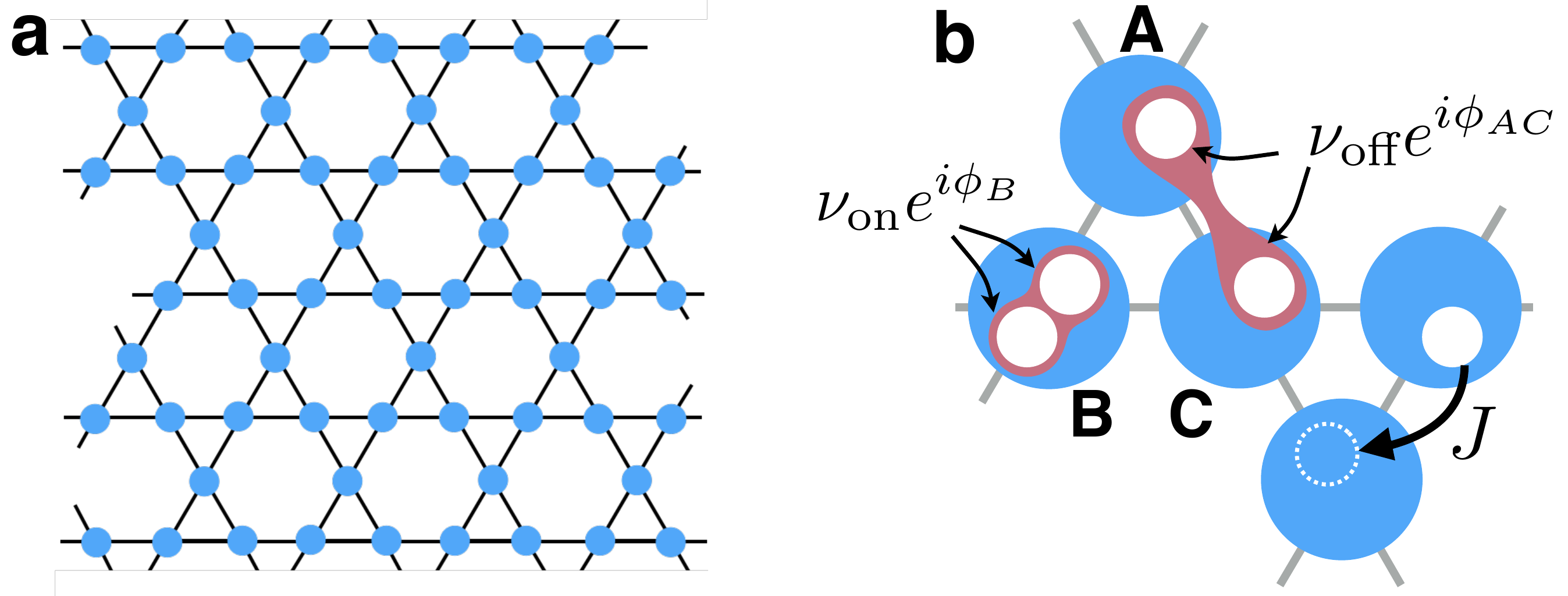}

	\caption{Setup figure: (a) An array of nonlinear cavities forming a kagome lattice. (b) Photons hop between  nearest-neighbor sites  with  rate $J$.  Each cavity is driven parametrically leading to the creation of  photon pairs on the same lattice site (rate $\nu_{\rm on}$) and on nearest-neighbor sites (rate $\nu_{\rm off}$). A spatial  pattern of  the  driving phase is imprinted on the parametric interactions, breaking the time-reversal symmetry (but preserving the  ${\cal C}3$ rotational symmetry).} 
	\label{fig:setup}
\end{figure}

Given these differences, it is natural to ask how anomalous "pairing" terms can directly lead to topological bosonic phases.  In this work, we study the topological properties of 2D systems described by Eq.~(\ref{eq:GenericH}), in the case where the underlying particle-conserving band structure has no topological structure, and where the parametric driving terms do not make the system unstable.  We show that the introduction of  particle non-conserving terms can break time-reversal symmetry (TRS) in a manner that is distinct from having introduced a synthetic gauge field, and can lead to the formation of bands having a non-trivial pattern of (suitably defined) quantized Chern numbers.  
This in turn leads to the formation of protected chiral edge modes: unlike the particle conserving case, these modes can mediate a protected inelastic scattering mechanism along the edge (i.e.~a probe field injected into the edge of the sample will travel along the edge, but emerge at a different frequency).  In general, the topological phases we find here are distinct both from those obtained in the particle-conserving case, and from those found in topological superconductors.  
We also discuss possible realizations of this model using a nonlinear photonic crystal or superconducting microwave circuits.

Note that recent works have explored topological features of bosonic quasiparticles in condensed phases; interactions with the condensate treated at the mean-field level yield a Hamiltonian having the general form of Eq.~(\ref{eq:GenericH}). These include a study of a magnonic crystal \cite{Shindou2013}, as well as general Bose-Einstein condensates in 1D \cite{Brandes2015}  and in 2D \cite{Liew2015}. Unlike our work, these studies did not explore the general role that tunable squeezing terms play in 
yielding topological states.  Further, in our case Eq.~(1) describes the "real" particles of our system, not quasiparticles defined above some background; this makes detection and potential applications much easier.   Most importantly, the topologically-protected inelastic scattering we describe is absent in those different settings.

\section{Results}

\subsection{Kagome lattice model}

For concreteness, we start with a system of bosons on a kagome lattice (see Fig.\ref{fig:setup}),
\begin{equation}
	\hat{H}_{0}=
	\sum_{\mathbf{j}} \omega_0 \hat{a}_{\mathbf{j}}^{\dagger}\hat{a}_{\mathbf{j}}-J 
	\sum_{\langle\mathbf{j,j'}\rangle}\hat{a}_{\mathbf{j}}^{\dagger}\hat{a}_{\mathbf{j'}}
	\label{eq:unperturbedHamiltonian}
\end{equation}
(we set $\hbar=1$). Here, we denote by $\hat{a}_{\mathbf{j}}$  the photon annihilation operator  associated with lattice site $\mathbf{j}$, where the 
vector site index has the form $\mathbf{j}=(j_{1},j_{2},s)$.  $j_{1},j_{2}\in Z$ labels a particular unit cell of the lattice, while the index $s=A,B,C$ labels the element of the sublattice.  $\langle\mathbf{j,j'}\rangle$ indicates the sum over nearest neighbors,
and $J$ is the (real valued) nearest neighbor hopping rate; $\omega_0$ plays the role of an on-site energy.  As there are no phases associated with the hopping terms, this Hamiltonian is time-reversal symmetric and topologically trivial.  We chose the kagome lattice because it is directly realizable both in quantum optomechanics \cite{Peano2015} and in arrays of superconducting cavity arrays \cite{Koch2010,Petrescu2012}; it is also the simplest model where purely local parametric driving can result in a topological phase.  


We next introduce quadratic squeezing terms to this Hamiltonian that preserve the translational symmetry of the lattice and that are no more non-local than our original, nearest-neighbor hopping Hamiltonian:
\begin{equation}
	\hat{H}_{L}=
		-\frac{1}{2}\left[
			\von \sum_{\mathbf{j}} e^{i \phi_s }
			\hat{a}_{\mathbf{j}}^{\dagger}\hat{a}_{\mathbf{j}}^{\dagger} 
			+ \voff 
			\sum_{\langle\mathbf{j,j'}\rangle}
			e^{i \phi_{ss'}}
			\hat{a}^{\dagger}_{\mathbf{j}}  \hat{a}^{\dagger}_{\mathbf{j'}}						
			\right] + h.c.
	\label{eq:parametricHamiltonian}
\end{equation}
Such terms generically arise from having a nonlinear interaction with a driven auxiliary pump mode on each site (which can be treated classically).  As we discuss below, the variation in phases in $\hat{H}_{L}$ from site to site could be achieved by a corresponding variation of the driving phase of the pump.  Note that 
we are working in a rotating frame where this interaction is time-independent, and thus $\omega_0$ should be interpreted as the detuning
between the parametric driving and the true on-site (cavity) frequency $\omega_{\rm cav}$ (i.e.~$\omega_0 = \omega_{\rm cav} - \omega_L/2$, where the parametric driving is at a frequency $\omega_L$). The parametric driving can cause the system to become unstable; we will thus require that the on-site energy $\omega_0$ is always sufficiently large that each parametric driving term is sufficiently non-resonant to ensure stability.

For a generic choice of phases in the parametric driving Hamiltonian of Eq.~(\ref{eq:parametricHamiltonian}), it is no longer possible to find a gauge where $\hat{H} = \hat{H}_0 + \hat{H}_L$ is purely real when expressed in terms of real-space annihilation operators:  hence, even though the hopping Hamiltonian $\hat{H}_0$ corresponds to strictly zero flux, {\it the parametric driving can itself break TRS}.  
In what follows, we will focus for simplicity on situations where time-reversal and particle-conservation are the only symmetries broken by the parametric driving:  they will maintain the inversion and $\mathcal{C}3$ rotational symmetry of the kagome lattice.  We will also make a global gauge transformation so that $\voff$ is purely real, while $\von = |\von| e^{i \varphi_\nu}$.  In this case, the only possible choices for the $\phi$ phases have the form $(\phi_A,\phi_B,\phi_C) = (\phi_{AB},\phi_{BC},\phi_{CA}) = \pm (0, \delta, 2 \delta)$ with 
$\delta =  2 \pi m_\nu / 3$, $m_\nu$ is an integer and is the vorticity of the parametric pumps.

%

\begin{figure*}
\includegraphics[width=1.5\columnwidth]{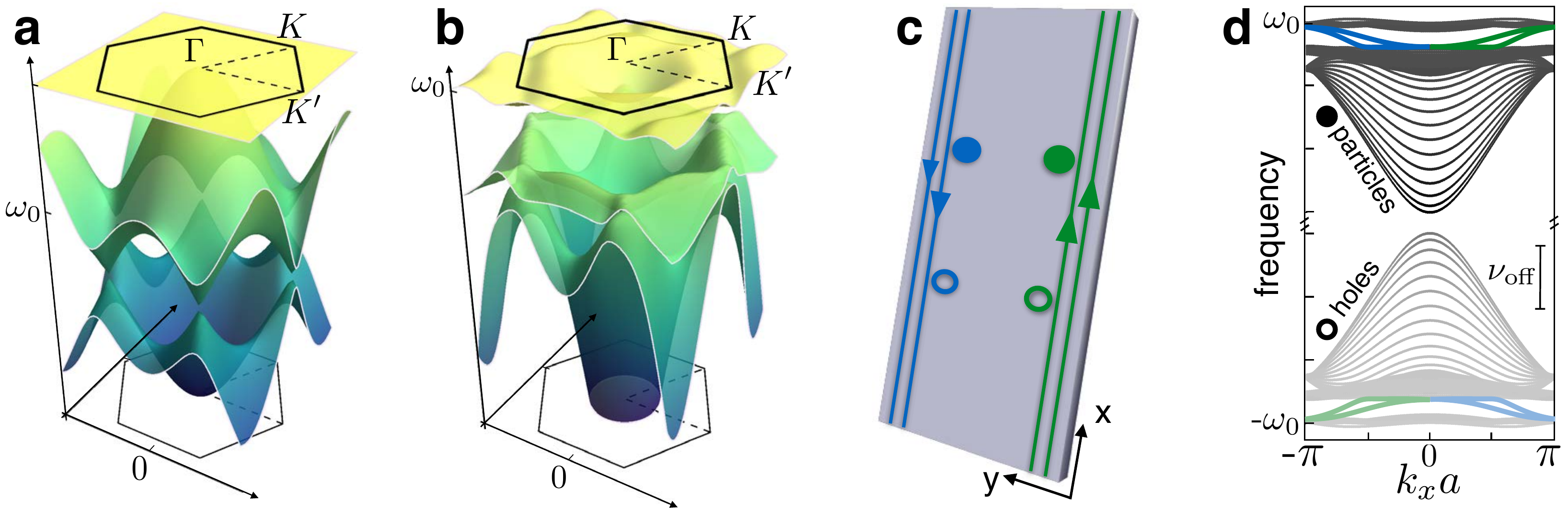}

\caption{Band structure: (a-b) 3D plots of the bulk band structure for $J=0.02\omega_0$. The hexagonal Brillouin zone is also shown. (a)  In the absence of parametric driving, neighboring bands touch at the rotational symmetry points $K$ and $K'$ and $\Gamma$. (b) The parametric driving opens a gap between subsequent bands. For the chosen parameter ($\nu_{\rm on}=-0.085$ and $\nu_{\rm off}=0.22$) there is a global band gap between the second and third band. (d) Hole and particle bands, $\pm E_m[k_x]$, in a strip geometry [sketched in (c)]. The line intensity is proportional to the weight of the corresponding resonance in the photon spectral function, see Appendix \ref{Appendix:Transport}.
The edge states localized on the right (left) edge, plotted in green (blue),  have positive (negative) velocity. }
	\label{fig:bandstructure}
\end{figure*}

\subsection{Gap opening and non-trivial topology}
$\hat{H}_0$ is the standard tight-binding kagome Hamiltonian for zero magnetic field, and does not have band gaps:
the upper and middle bands touch at the symmetry point $\Gamma\equiv(0,0)$, whereas the middle and lower
bands touch at the symmetry points $K=(2\pi/3,0)$ and $K'=(\pi/3,\pi/\sqrt{3})$ where they form Dirac cones [see Fig.~\ref{fig:bandstructure}(a)].

Turning on the pairing terms, the Hamiltonian $\hat{H} = \hat{H}_0 + \hat{H}_L$ can be diagonalized in the standard manner 
as $\hat{H} = \sum_{n,\bk} E_n[\bk] \hat{\beta}_{n,\bk}^\dagger \hat{\beta}^{\pd}_{n,\bk}$, where
the $\hat{\beta}_{n,\bk}$ are canonical bosonic annihilation operators determined by a Bogoliubov transformation of the form (see Appendix \ref{Appendix:Bogoliubovtrafo} for details):
\begin{equation}
	\hat{\beta}^\dagger_{n,\bk}
		=  \sum_{s=A,B,C} u_{n,\bk}[s]\hat{a}^\dagger_{\mathbf{k},s}
			-v_{n,\bk}[s] \hat{a}_{-\mathbf{k},s}.
		\label{eq:bogoliubov ansatz}
\end{equation}
Here, $\hat{a}_{\mathbf{k},s}$ are the annihilation operators in quasi-momentum space, and
 $n = 1,2,3$ is a band index; we count the bands by increasing energy. 
The photonic single-particle spectral function now shows resonances at both positive and negative frequencies, $\pm E_n[k]$, corresponding to ``particle''- and ``hole''-type bands, see Fig.~\ref{fig:bandstructure}(d).
 Because of the TRS breaking induced by the squeezing terms, the band structure described by $E_{n}[\bk]$ now exhibits gaps, see  Fig.~\ref{fig:bandstructure}(b); further, for a finite sized system, one also finds edge modes in the gap, see Fig.~\ref{fig:bandstructure}(d).  

The above behaviour suggests that the parametric terms have induced a non-trivial topological structure in the wavefunctions of the band eigenstates.  To quantify this, we first need to properly identify the Berry phase associated with a bosonic 
band eigenstate in the presence of particle non-conserving terms.
For each $\mathbf{k}$, the Bloch Hamiltonian $\hat{H}_{\rm \bk}$ corresponds to the Hamiltonian of a multi-mode parametric amplifier.  Unlike the particle-conserving case, the ground state of such a Hamiltonian is a multi-mode squeezed state with non-zero photon number; it can thus have a non-trivial Berry's phase associated with it when $\mathbf{k}$ is varied, see Appendix \ref{Appendix:BerryPhase}.  The Berry phase of interest for us will be the difference of this ground state Berry phase and that associated with a single quasiparticle excitation.  One finds that the resulting Berry connection takes the form 
\begin{equation}
{\cal A}_{n}=i\langle\mathbf{k},n | \hat{\sigma}_{z}\boldsymbol{\nabla_{k}} | \mathbf{k},n \rangle,
\end{equation}
Here,  the $6$-vector of Bogoliubov coefficients $|\mathbf{k}, n \rangle\equiv(u_{n,\bk}[A],u_{n,\bk}[B],u_{n,\bk}[C],v_{n,\bk}[A],v_{n,\bk}[B],v_{n,\bk}[C])$ plays the role of a singe-particle wavefunction,  and $\hat{\sigma}_z$ acts in the 'particle-hole-space', associating $+1$ to the $u$-components and $-1$ to the $v$-components,  see Appendix \ref{Appendix:BerryPhase} for details. 
These effective wavefunctions obey the symplectic normalization condition
\begin{align}
\langle  \mathbf{k},n |  \hat{\sigma}_{z}  |  \mathbf{k}',n'  \rangle
	& =  \sum_{s}
	 	 u_{n,\bk}^{*}[s]  u_{n',\bk'}[s]  -
	 	 v_{n,\bk}^{*}[s]  v_{n',\bk'}[s]  
		 \nonumber \\
	& = \delta_{\bk,\bk'} \delta_{n,n'}
	\label{eq:orthonormality}
\end{align}

Having identified the appropriate Berry connection for a band eigenstate, 
the Chern number for a band $n$ 
is then defined in the usual manner:
\begin{equation}
C_{n}=\frac{1}{2\pi}\int_{BZ}
\left( \boldsymbol{\nabla}\times{\cal A}_{n} \right) \cdot \hat{z} \label{eq:Chernnumber}
\end{equation}
This definition agrees with that presented in Ref.~\cite{Shindou2013} and (in 1D) Ref.~\cite{Brandes2015}; standard arguments
\cite{Shindou2013} show that the $C_n$ are integers with
the usual properties.  We note that, as for superconductors, breaking the $U(1)$  (particle-conservation) symmetry remains compatible with a first-quantized picture after doubling  the number of bands. The additional 'hole' bands  are connected to the standard 'particle' bands  by a particle-hole symmetry;  see Appendix \ref{Appendix:Bogoliubovtrafo}.  In bosonic systems, the requirement of stability implies that particle and hole bands can not touch. Thus,
the sum of the Chern numbers over the particle bands (with $E>0$) must be zero, and  there cannot be any edge states with energies below the lowest particle bulk band (or in particular, at zero energy); see Appendix \ref{Appendix:BerryPhase}.


In the special case where we only have onsite parametric driving (i.e.~$\nu_{\rm off}=0,\nu_{\rm on}\neq 0$), the Chern numbers can be calculated
analytically (see Appendix \ref{Appendix:phasediagram}).   They  are uniquely fixed by the pump vorticity.  If $m_\nu=0$, we have TRS and the band structure is gapless, while for $m_\nu=\pm 1$, $\vec{C} = (\mp 1, 0, \pm 1)$.  This set of  topological phases also occurs in a particle-number conserving model on the kagome lattice with a staggered magnetic field, i.e.~the Oghushi-Murakami-Nagaosa (OMN) model of the anomalous quantum Hall effect \cite{Ohgushi2000,Green2010}. 

\begin{figure}
	\includegraphics[width=0.9\columnwidth]{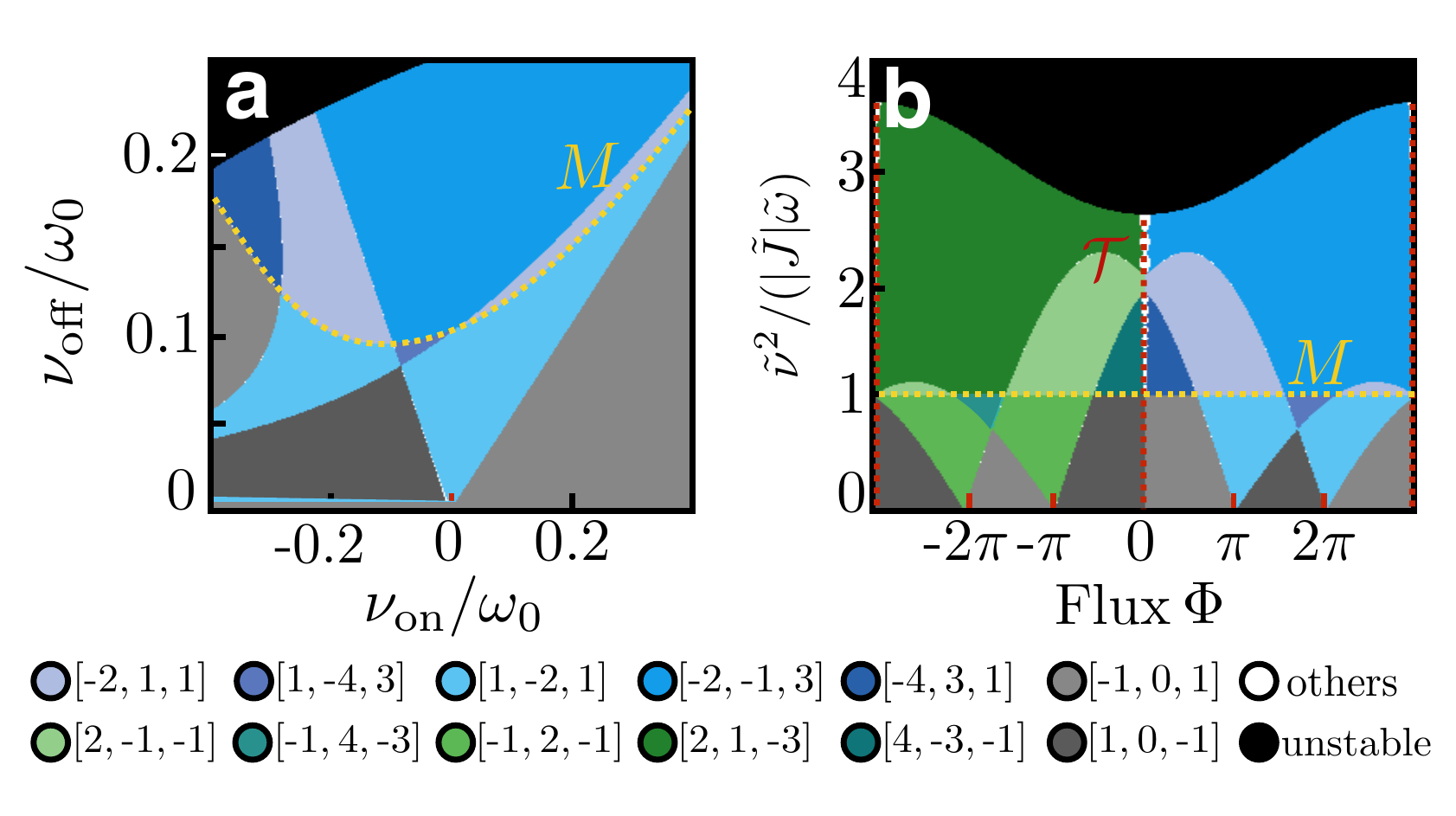}
	\caption{(a)Topological phase diagram for the parametrically driven kagome lattice model.  The $y$ ($x$) axis corresponds to the strength of the onsite parametric drive $\von$ (off-site parametric drive $\voff$), and different colours correspond to different triplets $\vec{C} = (C_1, C_2, C_3)$ of Chern numbers for the three bands of the model.
	Note that  only the gray and dark-gray phases are found in the particle-conserving version of our model with a staggered field.
	  We have fixed the hopping rate $J/\omega_0 = 0.02$, and the vorticity of the pump $m_\nu=1$.
	(b)  Same phase diagram, but now plotted in terms of the effective flux $\tilde{\Phi}$ and effective parametric drive $\tilde{\nu}$ experienced by $\alpha$ quasiparticles.  See text for details.} 
	\label{fig:PhaseDiagram}
\end{figure}

In the general case, including offsite parametric driving, entirely new phases appear. We have computed the Chern numbers for that case numerically, using the approach of Ref.~\cite{Fukui2005}.  In Fig.~\ref{fig:PhaseDiagram}(a), we show the topological phase diagram of our system, where $J/\omega_0$ and $m_\nu$ 
are held fixed, while the parametric drive strengths $\von,\voff$ are varied.   Different colors correspond to different triplets $\vec{C} \equiv (C_1, C_2, C_3)$ of the band Chern numbers, with gray and dark-gray corresponding to the two phases already present in the OMN model.  
Strikingly, a finite  off-diagonal coupling $\nu_{\rm off}$  generates a 
large variety of phases which are not present in the OMN model. The border between different topological
phases represent topological phase transitions, and correspond to parameter values where a pair of bands touch at a particular
symmetry point; we discuss this further below.  

\subsection{Effective model}  

To gain further insight into the structure of the topological phases found above, it is useful to work in a ``dressed-state" basis that eliminates the local parametric driving terms from our Hamiltonian.  We thus
first diagonalize the purely local terms in the Hamiltonian; for each lattice site $\bj$ we have
\begin{eqnarray}  
 	\hat{H}_{\bj} & = & 
	\omega_0 \hat{a}^\dagger_{\bj} \hat{a}^{\pd}_{\bj} 
	-\frac{1}{2}\left[
			\von  e^{i \phi_\bj }
			\hat{a}_{\mathbf{j}}^{\dagger}\hat{a}_{\mathbf{j}}^{\dagger}  + h.c. \right]
			= \tilde{\omega} \halpha^\dagger_{\bj} \halpha^{\pd}_{\bj}.
			\label{eq:HNew}
\end{eqnarray}
Here $\tilde{\omega} = \sqrt{\omega_0^2 - \von^2}$, and the annihilation operators $\halpha_{\bj}$ are given by a local Bogoliubov (squeezing) transformation 
$\halpha_{\bj} = e^{i \phi_\bj}e^{-i\varphi_\nu/2} \left( \cosh r a_{\bj} -  e^{i \phi_\bj}e^{i\varphi_\nu} \sinh r \hat{a}^\dagger_{\bj} \right)$, where the squeezing factor $r$ is
\begin{equation}
r=\frac{1}{4}\ln\left[\frac{\omega_0+\nu_{{\rm on}}}{\omega_0-\nu_{{\rm on}}}\right].\label{eq:squeezingfactor}
\end{equation}
On a physical level, the local parametric driving terms attempt to drive each site into a squeezed vacuum state with squeeze parameter $r$; the $\hat{\alpha}_\bj$ quasiparticles correspond to excitations above this reference state.  Note that we have included an overall phase factor in the definition of the $\halpha_{\bj}$ which will simplify the final form of the full Hamiltonian.

In this new basis of local quasiparticles, our full Hamiltonian takes the form
\begin{eqnarray}
	\hat{H}=\sum_{\mathbf{j}} 
		\tilde{\omega} \hat{\alpha}_{\mathbf{j}}^{\dagger}\hat{\alpha}_{\mathbf{j}}
	-\sum_{\langle\mathbf{j,l}\rangle} 
		\tilde{J}_{\mathbf{jl}} 
		\hat{\alpha}_{\mathbf{j}}^{\dagger}\hat{\alpha}_{\mathbf{l}}
	-\left( \frac{\tilde{\nu} }{2}  \sum_{\langle\mathbf{j,l}\rangle}
	\hat{\alpha}_{\mathbf{j}}^{\dagger} \hat{\alpha}_{\mathbf{l}}^{\dagger}+h.c. \right). \nonumber \\
\end{eqnarray}
The transformation has mixed the hopping terms with the non-local parametric terms: The effective counter-clockwise hopping matrix element is
\begin{align}
	\tilde{J}_{\mathbf{jl}} & = 
		J e^{i \delta} + \nonumber\\
			&  \!\!\!\!e^{3i \delta / 2} \left[
			2 J \cos \Big(\frac{\delta}{2} \Big) \sinh^2r  +
			\voff  \sinh 2r \cos \Big(\frac{\delta}{2}+\varphi_\nu \Big) \right], \label{eq:Jeff}
\end{align}
and the magnitude of the effective non-local parametric driving is
\begin{align}
	|\tilde{\nu}|  = 
		 \big|& 
			\voff  e^{-i(\delta/2+\varphi_\nu)}  + 
			  2\voff\cos(\delta/2+\varphi_\nu)  \sinh^2 r\nonumber\\
		&+ J \sinh 2r\cos(\delta/2) \big|.
\end{align}
Note that the phase of $\tilde{\nu}$ can be eliminated by a global gauge transformation, and hence it plays no role; we thus take $\tilde{\nu}$ to be real in what follows.

Our model takes on a much simpler form in the new basis:  the onsite parametric driving is gone, and the non-local parametric driving is real.  Most crucially, the effective hoppings now can have spatially-varying phases, which depend both on the vorticity of the parametric driving in $\hat{H}_L$  (through $\delta$), and  the magnitude of the on-site squeezing (through $r$).  In this transformed basis, the effective hopping phases are the only route to breaking TRS. If the parametric terms were not present, the complex phases would correspond in the usual manner to a synthetic gauge field 
(i.e.~the effective flux $\Phi$ piercing a triangular plaquette would be $\Phi=3\arg \tilde{J}$). 
Our model has thus been mapped on to the standard OMN model for the anomalous quantum Hall effect, with an additional (purely real) nearest-neighbour two-mode squeezing interaction. However, we note that strictly speaking  $\Phi$ can not be interpreted as a flux in the presence of the additional parametric terms: a 'flux' of $2\pi$ can not be anymore eliminated by a gauge transformation because the complex phases 
reappear in the parametric terms. In that case, only a periodicity of $6\pi$ in $\Phi$ is retained, since that corresponds to having trivial hopping phases of $2\pi$.

Understanding the topological structure of this transformed Hamiltonian is completely sufficient for our purposes:  one can easily show that the Chern number of a band is invariant under any local Bogoliubov transformation, hence the Chern numbers obtained from the transformed Hamiltonian in 
Eq.~(\ref{eq:HNew}) will coincide exactly with those obtained from the original Hamiltonian in Eq.~(\ref{eq:parametricHamiltonian}).  We thus see that the topological structure of our system is controlled completely through only three dimensionless parameters:  the 'flux' $\Phi$ associated with the hopping phase, the ratio $|\tilde{\nu} / \tilde{J}|$ and the ratio $\tilde{\omega} / |\tilde{J}|$.  

The topological phase diagram for the effective model is shown in Fig.\ \ref{fig:PhaseDiagram}(b).
Again, one sees that as soon as the effective non-local parametric drive $\tilde{\nu}$ is non-zero, topological phases distinct from the standard (particle-conserving) OMN model are possible. The sign of the parametric pump vorticity $m_\nu$ determines the sign of the effective flux $\Phi$, c.f.~~Eq.\ \ref{eq:Jeff}.  As such, 
the right half of Fig. \ref{fig:PhaseDiagram}(b) (corresponding to $\Phi > 0$) is a deformed version of the phase diagram of the original model for pump vorticity $m_\nu = 1$, as plotted in Fig.\ \ref{fig:PhaseDiagram}(a).  Changing the sign of $m_\nu$ (and hence $\Phi$) simply flips the sign of all Chern numbers, see Appendix \ref{Appendix:phasediagram}.

Our effective model provides a more direct means for understanding the boundaries between different topological phases.  Most of these are associated
with the crossing of bands at one or more high-symmetry points in the Brillouin zone; this allows an analytic calculation of the phase boundary (see Appendix \ref{Appendix:phasediagram}).  Perhaps most striking in Fig.~\ref{fig:PhaseDiagram}(b) is the horizontal boundary (labelled $\mathcal{M}$), occurring at a finite value of the effective offsite parametric drive, $\tilde{\nu} \approx \sqrt{\tilde{J} \tilde{\omega}}$.
This boundary is set by the closing of a band gap at the $M$ points; as these points are associated with the decoupling of one sublattice from the other two, this boundary is insensitive to the flux $\Phi$.  Similarly, the vertical line labeled $\mathcal{T}$ denotes a line where the system has TRS, and all bands cross at 
the symmetry points $K$, $K'$ and $\Gamma$.  The case of zero pump vorticity $m_\nu = 0$ (not shown) is also interesting.  Here, the effective flux $\Phi$ depends on the strength of the parametric drivings, but is always constrained to be $0$ or $3 \pi$.  This implies that the effective Hamiltonian has TRS, even though the original Hamiltonian may not (i.e.~ if $\textrm{Im } \voff \neq 0$, the original Hamiltonian does not have TRS).  For $m_{\nu}=0$,  the parametric drivings do not open any band gap and the Chern numbers are not well defined.



\subsection{Edge states and transport} 

Despite their modified definition, the Chern numbers associated with our Bogoliubov bands still guarantee the existence of protected chiral edge modes in a system with boundaries via a standard bulk-boundary correspondence, see Appendix \ref{Appendix:Bulk/Boundarycorrespondence}.  These states can be used to transport photons, by exciting them with an auxiliary probe laser beam which is focused on an edge site and at the correct frequency. The lack of particle-number conservation manifests itself directly in the properties of the edge states:  along with the standard  elastic transmission they can also mediate  {\it inelastic} scattering processes. In terms of the original lab frame, light injected at a frequency $\omega_{p} $ can emerge on the edge at frequency  $\omega_L-\omega_{p}$ where $\omega_L$ is the frequency of the laser   parametrically driving the system. This is analogous to  the idler output of a parametric amplifier. Here, both signal and idler  have a topologically protected chirality.

\begin{figure}
	\includegraphics[width=0.94\columnwidth]{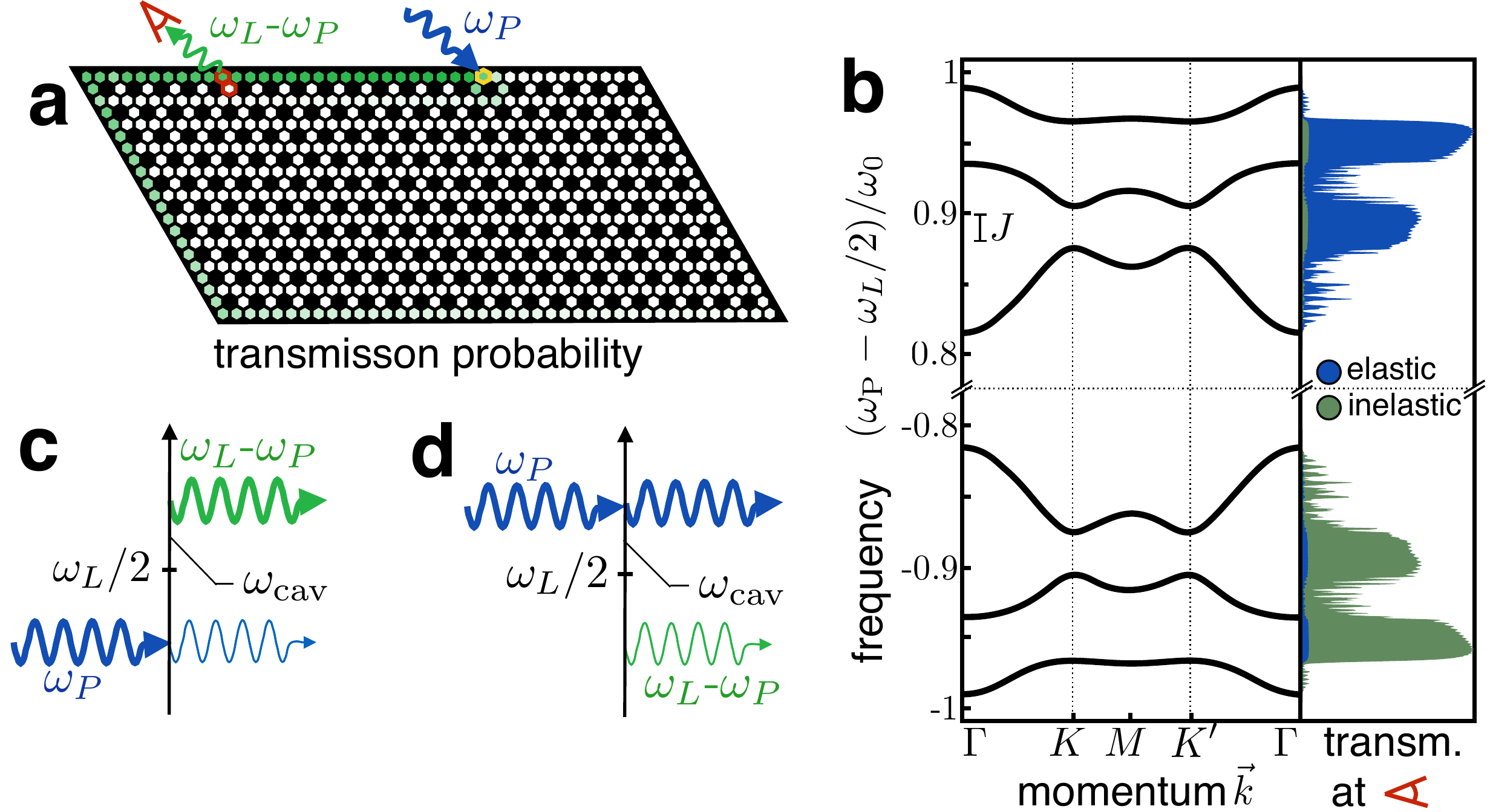}
	\caption{Topologically protected transport in a finite system: (a) A probe beam  at frequency $\omega_p$ inside  the bulk band gap is focused on the edge of a finite sample. The probability map  of the light transmitted inelastically at frequency $\omega_L-\omega_p$ clearly shows that the transport is chiral. (b) The elastic and inelastic transmission probability to a pair of sites along the edges [indicated in red in (a)] is plotted in blue and green, respectively. A cut through the bulk bands is shown to the left. (c,d) Sketch of the relevant scattering processes and energy scales. The inelastic (elastic) transmission has a larger rate when the light is injected in the {\it hole} ({\it particle}) band gap.  Parameters: $J/\omega_0=0.02$, $\nu_{\rm on}/\omega_0=0.4$, $\nu_{\rm off}/\omega_0=0.02$, $\kappa/\omega_0=0.001$. In (a) $\omega_p-\omega_L/2=0.95\omega_0$ \label{Fig:transport}} 
\end{figure}

Shown in Fig.~\ref{Fig:transport} are the results of a linear response calculation describing such an experiment, applied to a finite system with corners. We incorporate a finite photon decay rate $\kappa$ in the standard input-output formalism, see Appendix \ref{Appendix:Transport} for details.  Narrow-band probe light  inside a topological band gap is applied to a site on the edge, and the resulting inelastic transmission probabilities to each site on the lattice are plotted, see  Fig.~\ref{Fig:transport}(a). One clearly sees that the probe light is transmitted in a uni-directional way along the edge of the sample, and is even able to turn the corner without significant backscatter. The corresponding elastic transmission [not shown] is also chiral and shows the same spatial dependence. In Fig.~\ref{Fig:transport}(b) we show the  elastic and inelastic transmittions  to the sites  indicated in red  (rescaled by the overall transmission, $1-R$ where $R$ is the reflection probability at the injection site) as a function of the probe frequency $\omega_p$. By scanning the laser probe frequency one can separately address {\it particle} and {\it hole} band gaps. The relative intensity of the inelastic scattering component is highly enhanced  when the probe beam is inside a {\it hole} band gap, see also the sketches in Fig.~\ref{Fig:transport}(c-d). 
 When the parametric interaction between the $\hat{\alpha}$ quasiparticles is negligible, the ratio of elastic and inelastic 
transmissions  depends only on the squeezing factor $r$, [c.~f.~Eq.~(\ref{eq:squeezingfactor})], see Appendix \ref{Appendix:Transport}.

\subsection{Physical realization} 

Systems of this type could be
implemented in 2D photonic crystal coupled cavity arrays fabricated
from nonlinear optical $\chi^{(2)}$ materials \cite{Yariv2002,Eggleton2011,Dahdah2011}. The array of optical
modes participating in the transport would be supplemented by pump
modes (resonant with the pump laser at twice the frequency). One type
of pump mode could be engineered to be spatially co-localized with
the transport modes ($\nu_{{\rm on}}$ processes), while others could
be located in-between ($\nu_{{\rm off}}$). The required periodic
phase pattern of the pump laser can be implemented using spatial light
modulators or a suitable superposition of several laser beams impinging
on the plane of the crystal. Optomechanical systems offer another
route towards generating optical squeezing terms \cite{Painter2013,Regal2013},
via the mechanically induced Kerr interaction, and this could be exploited
to create an optomechanical array with a photon Hamiltonian of the
type discussed here. Alternatively, these systems can be driven by
two laser beams to create \emph{phononic} squeezing terms \cite{Kronwald2013}. A fourth
alternative consists in superconducting microwave circuits of coupled
resonators, where Josephson junctions can be embedded to introduce
$\chi^{(2)}$ and higher-order nonlinearities, as demonstrated in \cite{Bergeal2010,Abdo2013}.

\section{Conclusions and Outlook} 

In this work, we have shown how tunable squeezing interactions in a photonic system are flexible tools that allow the creation of new kinds of 
topological bosonic phases.  We further demonstrated that the protected edge channels supported by such phases allow uni-directional elastic and inelastic coherent photon transport.  Our work opens the door to a number of interesting new directions.  On the more practical side, one could attempt to exploit the unique edge states in our system to facilitate directional, quantum-limited amplification.  On the more fundamental level, one could use insights from the corresponding disorder problem \cite{Gurarie2003} and attempt to develop a full characterization of particle non-conserving bosonic topological states that are described by quadratic Hamiltonians.  This would then be a counterpart to the classification already developed for fermionic systems \cite{Ryu2010}.

%


VP, CB, and FM acknowledge support by an ERC Starting Grant OPTOMECH, by the DARPA project ORCHID, and by the European Marie-Curie ITN network cQOM. MH and AC acknowledge support from NSERC.  

\appendix
\section{Details of the calculation of the band structure}
\label{Appendix:Bogoliubovtrafo}
\subsection{Bogoliubov transformation and first-quantized picture}
Here, we show how to find the Bogoliubov transformation Eq.~(\ref{eq:bogoliubov ansatz}) which diagonalizes the Hamiltonian $\hat{H}$ defined
in Eqs.~(\ref{eq:unperturbedHamiltonian},\ref{eq:parametricHamiltonian}). It is convenient to switch to a first quantized picture
by casting the second quantized Hamiltonian $\hat{H}$ in the form
\[
\hat{H}=\frac{1}{2}\sum_{\mathbf{k}}\begin{pmatrix}\hat{\boldsymbol{a}}_{\mathbf{k}}^{\dagger}\\
\hat{\boldsymbol{a}}_{\mathbf{-k}}
\end{pmatrix}^{T}h(\mathbf{k)}\begin{pmatrix}\hat{\boldsymbol{a}}_{\mathbf{k}}\\
\hat{\boldsymbol{a}}_{\mathbf{-k}}^{\dagger}
\end{pmatrix}
\]
where $\hat{\boldsymbol{a}}_{\mathbf{k}}=(\hat{a}_{\mathbf{k}A},\hat{a}_{\mathbf{k}B},\hat{a}_{\mathbf{k}C})^{T}$
and $h(\mathbf{k})$ is the Bogoliubov de Gennes Hamiltonian. By plugging the Bogoliubov ansatz Eq.~(\ref{eq:bogoliubov ansatz}) into the Heisenberg equation of motion for the normal modes ladder operators
$\hat{\beta}^\dagger_{n,\mathbf{k}}$,
\[
 -i\frac{d}{dt}\hat{\beta}^\dagger_{n,\mathbf{k}}=[\hat{H},\hat{\beta}^\dagger_{n,\mathbf{k}}]=E_n[\mathbf{k}]\hat{\beta}^\dagger_{n,\mathbf{k}},
\]
one immediately finds the generalized eigenvalue problem
\begin{equation}
\hat{h}(\mathbf{k})|\mathbf{k}_{n}\rangle=E_{n}[\mathbf{k}]\hat{\sigma}_{z}|\mathbf{k}_{n}\rangle .\label{eq:eigenvalueproblem}
\end{equation}
Here, the matrix $\hat{\sigma}_{z}$ is (minus) the identity for the (hole) particle
sector,
\[
\hat{\sigma}_{z}=\begin{pmatrix}1\!\!1_{3} & 0\\
0 & -1\!\!1_{3}
\end{pmatrix}.
\]
The bosonic quadratic Hamiltonian $\hat{H}$ describes the dynamics linearized around a  \emph{stable classical solution} if it is possible to find
its  normal mode decomposition.  In other words, for all $\mathbf{k}$ the eigenvalue problem Eq.~(\ref{eq:eigenvalueproblem}) should have a set of
three orthonormal (particle-like) solutions, c.~f.~Eq.~(\ref{eq:orthonormality}). A sufficient and necessary criterion for stability is that
 all eigenvalues of $\hat{\sigma}_{z}\hat{h}(\mathbf{k})$ are real, see discussion below.

\subsection{Particle-hole symmetry}
The symplectic eigenvalue problem (\ref{eq:eigenvalueproblem})
has an embedded particle-hole symmetry: for any  particle-like
solution with momentum $\mathbf{k}$, energy $E_{n}[\mathbf{k}]$,
and wavefunction $|\mathbf{k}_{n}\rangle$ there is a
hole-like solution with momentum $-\mathbf{k}$, 
energy $-E_{n}[\mathbf{k}]$, and wavefunction
\[
|-\mathbf{k}_{-n}\rangle=\sigma_{x}{\cal K}|\mathbf{k}_{n}\rangle\equiv{\cal C}|\mathbf{k}_{n}\rangle.
\]
(notice that the hole-like solutions have negative length $\langle\mathbf{k}_{-n}|\hat{\sigma}_z|\mathbf{k}_{-n}\rangle=-1$).
Here, operator ${\cal K}$ gives the complex conjugate of the wavefunction
and $\hat{\sigma}_{x}$ exchanges particles and holes, 
\[
\hat{\sigma}_{x}=\begin{pmatrix}0 & 1\!\!1_{3}\\
1\!\!1_{3} & 0
\end{pmatrix}.
\]
In other words, the Bogoliubov de Gennes Hamiltonian has the generalized
symmetry ${\cal {\cal C}^{\dagger}}\hat{h}(\mathbf{k}){\cal C}=-\hat{h}(-\mathbf{k})$
where the charge conjugation operator ${\cal C}$ is anti-unitary
and ${\cal C}^{2}=1\!\!1_{6}$. Thus, our system represents the Bosonic
analogue of a superconductor in the Class $D$ of the standard topological
classification.

\subsection{Numerical calculation of the band structure}

From Eqs.~(\ref{eq:unperturbedHamiltonian},\ref{eq:parametricHamiltonian}) one can immediately derive the explicit
expression of the Bogoliubov de Gennes Hamiltonian
\begin{equation}
\hat{h}(\mathbf{k})=\begin{pmatrix}\omega_{0}-J\hat{\tau}(\mathbf{k}) & h_{L}(\mathbf{k)}\\
h_{L}^{\dagger}(\mathbf{k}) & \omega_{0}-J\hat{\tau}(\mathbf{k})
\end{pmatrix}\label{eq:BogoliubovDeGennes-2}
\end{equation}
where $\hat{h}_{L}=-\nu_{{\rm on}}\hat{\Lambda}^{m_{\nu}}-\nu'_{{\rm off}}(\hat{\Lambda}^{m_{\nu}}\hat{\tau}+\hat{\tau}\hat{\Lambda}^{m_{\nu}})$ ($m_\nu$ is the pump vorticity).
The matrices $\hat{\tau}(\mathbf{k})$ and $\hat{\Lambda}$ are single-particle
operators acting on the sub-lattice degrees of freedom. They have
matrix elements: $\Lambda_{AA}=1$, $\Lambda_{BB}=e^{i2\pi/3},$ $\Lambda_{CC}=e^{-2i\pi/3},$
$\tau_{BA}=\tau_{AB}^{*}=1+e^{i\mathbf{k}\cdot\boldsymbol{a}_{1}},$
$\tau_{CB}=\tau_{BC}^{*}=1+e^{i\mathbf{k}\cdot\boldsymbol{a}_{2}},$
$\tau_{AC}=\tau_{CA}^{*}=1+e^{i\mathbf{k}\cdot\boldsymbol{a}_{3}}$
 where $\boldsymbol{a}_{1}=(-1,-\sqrt{3})$, $\boldsymbol{a}_{2}=(2,0)$,
and $\boldsymbol{a}_{3}=(-1,\sqrt{3})$ are lattice vectors. All other
matrix elements are zero. Notice that $\hat{\Lambda}$ raises the
quasi-angular momentum by one unit: $\hat{\Lambda}|p,m\rangle=|p,m+1\rangle$
where $|p,m\rangle=(1,\exp[i2m\pi/3],\exp[-i2m\pi/3],0,0,0)$ and
likewise for the holes. Here, $m=0$ is the vortex free eigenstate,
$m=1$ has a vortex, and $m=-1$ an anti-vortex ($m$ is defined modulo
$2$). Thus, a hole with quasimomentum $m$ is converted into a particle
with quasi-angular momentum $m+m_{\nu}$. In other words, a pair of
down-converted photons have quasi-angular momenta $-m$ and $m_{\nu}+m$,
respectively. The additional quasimomentum $m_{\nu}$ is provided
by the pump photons. In order to write the off-diagonal parametric interaction compactly in terms of the quasi-angular momentum raising operator $\hat{\Lambda}$  we have introduced the rescaled off-diagonal parametric coupling $\voff'=e^{-i\delta/2}\voff$ for pump vorticity $m_\nu=\pm 1$ and $\voff'=\voff/2$ for pump vorticity $m_\nu=0$.

For the effective model Eq.~(\ref{eq:HNew}) the Bogoliubov de
Gennes Hamiltonian reads 
\begin{equation}
\hat{\tilde{h}}(\mathbf{k})=\begin{pmatrix}\tilde{\omega}-\tilde{J}\hat{\tilde{\tau}}(\mathbf{k},\Phi) & -\tilde{\nu}\hat{\tau}(\mathbf{k})\\
-\tilde{\nu}\hat{\tau}(\mathbf{k}) & \tilde{\omega}-\tilde{J}\hat{\tilde{\tau}}(\mathbf{k},-\Phi)
\end{pmatrix}.\label{eq:BogoliubovDeGenneseff}
\end{equation}
Here, we have introduced the hopping matrix $\hat{\tau}(\mathbf{k},\Phi)$
in the presence of the synthetic magnetic field flux $\Phi$. It has
the following non zero matrix elements: $\tilde{\tau}_{BA}=\tilde{\tau}_{AB}^{*}=e^{i\Phi/3}(1+e^{i\mathbf{k}\cdot\boldsymbol{a}_{1}}),$
$\tilde{\tau}_{CB}=\tilde{\tau}_{BC}^{*}=e^{i\Phi/3}(1+e^{i\mathbf{k}\cdot\boldsymbol{a}_{2}}),$
$\tilde{\tau}_{AC}=\tilde{\tau}_{CA}^{*}=e^{i\Phi/3}(1+e^{i\mathbf{k}\cdot\boldsymbol{a}_{3}})$. 
In the most general case, we calculate the band structure and the single-particle wavefunctions  by diagonalizing the $6\times6$ matrix $\hat{\sigma}_{z}\hat{h}(\mathbf{k)}$
numerically.

\subsection{Analytical calculation of the band structure close to the symmetry points}
\label{Appendix:symmetrypoints}

One can gain much insight on the array dynamics, including the
stability requirements and  the array  topology, by calculating analytically the band strucure 
at the symmetry points. This is a particularly easy task at the rotational
symmetry points $\Gamma$, $K$, and $K'$. There, the hopping matrix
$\hat{\tau}$ is diagonal in the basis of the quasi-angular momentum
eigenstates. Thus, in this basis the Hamiltonian becomes block diagonal
with $2$-dimensional blocks. Each block is described by a two-mode
squeezing Hamiltonian, except for the quasi-momentum $\mathbf{k}=\Gamma$
and $m=-m_{\nu}$, when it is a single mode squeezing Hamiltonian
\begin{eqnarray*}
\hat{H}_{L}(\mathbf{k}) & = & -\sum_{m}[\nu_{{\rm on}}+\nu'_{{\rm off}}(\tau_{m}+\tau_{m-m_{\nu}})]\hat{\tilde{a}}_{\mathbf{\mathbf{k}},m}^{\dagger}\hat{\tilde{a}}_{\mathbf{-\mathbf{k}},m_{\nu}-m}^{\dagger}\\
 &  & +H.c.,\quad{\rm for}\quad\mathbf{k}=K,K',\\
\hat{H}_{L}(\Gamma) & = & -\big\{\sum_{m\neq m_{\nu}}[\nu_{{\rm on}}+\nu'_{{\rm off}}(\tau_{m}+\tau_{m-m_{\nu}})]\hat{\tilde{a}}_{\mathbf{\Gamma},m}^{\dagger}\hat{\tilde{a}}_{\mathbf{\Gamma},m_{\nu}-m}^{\dagger}\\
 &  & -(\nu_{{\rm on}}/2+\nu'_{{\rm off}}\tau_{m_{\nu}})\hat{\tilde{a}}_{\mathbf{\Gamma},-m_{\nu}}^{\dagger2}\big\}+H.c.
\end{eqnarray*}
where $\hat{\tilde{a}}_{\mathbf{\mathbf{k}},m}$ is the creation operator
of an excitation with quasimomentum $\mathbf{k}$ ($\mathbf{k}=\Gamma,K,K'$)
and quasi-angular momentum $m$. Moreover, $\tau_{m}$ indicates the
corresponding eigenvalue of the hopping matrix $\hat{\tau}$: $\tau_{0}(\Gamma)=4$,
$\tau_{1}(\Gamma)=\tau_{-1}(\Gamma)=\tau_{1}(K)=\tau_{-1}(K')=-2$,
$\tau_{0}(K)=\tau_{1}(K')=\tau_{0}(K')=\tau_{-1}(K)=1.$ By diagonalizing
the squeezing Hamiltonian we find the general expression for the eigenvalues
\begin{eqnarray}
E_{m} & = & -J\frac{\tau_{m}-\tau_{m-m_{\nu}}}{2}+\left[\left(\omega_{{\rm 0}}-J\frac{\tau_{m}+\tau_{m-m_{\nu}}}{2}\right)^{2}\right.\nonumber \\
 &  & \left.-\left|\nu_{{\rm on}}+\nu'_{{\rm off}}(\tau_{m}+\tau_{m-m_{\nu}})\right|^{2}\right]^{1/2}.\label{eq:spectrumrotsym}
\end{eqnarray}
From an analogous calculation we obtain the spectrum of the effective
model at the rotation symmetry points 
\begin{equation}
\tilde{E}_{m}=\left[\left(\tilde{\omega}_{{\rm }}-\tilde{J}\tilde{\tau}_{m}(\Phi)\right)^{2}-\tau_{m}^{2}\tilde{\nu}^{2}\right]^{1/2}\label{eq:spectrumsymeffective}
\end{equation}
where $\tilde{\tau}_{m}(\mathbf{k},\Phi)$ are the eigenvalues of
$\hat{\tilde{\tau}}$: $\tilde{\tau}_{0}(\Gamma)=4\cos[\Phi/3]$,
$\tau_{\pm1}(\Gamma)=4\cos[2\pi/3\mp\Phi/3]$, $\tau_{1}(K)=\tau_{-1}(K')=-2\cos[\Phi/3]$,
$\tau_{1}(K')=\tau_{0}(K)=2\cos[\pi/3-\Phi/3]$, $\tau_{-1}(K)=\tau_{0}(K')=2\cos[\pi/3+\Phi/3]$.

\subsection{Stability analysis}
\label{Appendix:stability}

The system is stable when all eigenvalues of $\hat{\sigma}_{z}\hat{h}(\mathbf{k})$
are real. If all eigenenergies of the unperturbed Hamiltonian $\hat{H}_{0}$
have the same sign the parametric interaction is off-resonant and
the system is stable if the parametric couplings are below a finite
threshold. On the contrary, if the unperturbed band touches the zero-energy
axis, the parametric interaction is resonant for the zero energy modes
leading to an instability for any arbitrarily small value of the coupling. Thus,
the parametric instability sets an upper limit to the hopping $J$.
For concreteness, we consider a positive onsite energy $\omega_{0}$
(corresponding to a red detuned drive). In this case, all eigenenergies
of $\hat{H}_{0}$ are positive if $J<\omega_{0}/4$. In this case,
the system is stable for sufficiently small values of the parametric
couplings $\nu_{{\rm on}}$ and $\nu_{{\rm off}}.$ Nevertheless,
the threshold of an instability is reached as soon as the lowest eigenenergy
of a particle-type band becomes zero for a finite value of the parametric
couplings $\nu_{{\rm on}}$ and $\nu_{{\rm off}}.$ For the parameters
of Fig.~\ref{fig:PhaseDiagram}  the lowest band touches the zero energy
axis at the $\Gamma$ point. Thus, we can find an analytical expression
for the instability threshold using the solutions at the rotational
symmetry points Eqs. (\ref{eq:spectrumrotsym},\ref{eq:spectrumsymeffective}).
In Fig.~3(a) the state with zero energy at the border of the unstable
region has zero quasimomentum and an anti-vortex. Thus, the instability threshold is given by setting $E_{-1}(\Gamma)=0$.
In Fig.~3(b) the state with zero energy has zero quasi-momentum and
quasi-angular momentum. By setting $\tilde{E}_{0}(\Gamma)=0$, we
find a simple expression for the instability threshold, $\tilde{\nu}=\tilde{\omega}/4-\tilde{J}\cos[\Phi/3].$ 

\section{Details of the definition and properties of the symplectic Chern number}
\label{Appendix:BerryPhase}
\subsection{Berry phase of a Bogoliubov quasi-particle}

For an excitation conserving Hamiltonian, the Chern number of the
$m$-th band can be viewed as a sum of Berry phases accumulated on
a set of closed loops covering the whole Brillouin zone. In this case,
the state which accumulates the relevant Berry phase is the $m$-th
eigenstate of the single-particle Hamiltonian $\hat{h}_{\mathbf{k}}$
(a block with quasimomentum $\mathbf{k}$ of the single-particle Hamiltonian
$\hat{h}$). Below, we show that one can naturally extend this definition
of the Chern number to any bosonic Hamiltonian including anomalous
terms by identifying the relevant Berry phase in a second-quantized
setting.

For each quasi-momentum $\mathbf{k}$, the second-quantized block
\[
\hat{H}_{\mathbf{k}}=\begin{pmatrix}\hat{\boldsymbol{a}}_{\mathbf{k}}^{\dagger}\\
\hat{\boldsymbol{a}}_{\mathbf{-k}}
\end{pmatrix}^{T}h(\mathbf{k)}\begin{pmatrix}\hat{\boldsymbol{a}}_{\mathbf{k}}\\
\hat{\boldsymbol{a}}_{\mathbf{-k}}^{\dagger}
\end{pmatrix}
\]
of the full bosonic Hamiltonian $\hat{H}$ is a six-mode squeezing
Hamiltonian. If we regard the quasi-momentum $\mathbf{k}$ as an external
parameter, we can ask ourself what is the additional Berry phase accumulated
by a single Bogoliubov quasi-particle in a specific band $n$ while
the quasi-momentum is varied adiabatically over a closed loop.  In
other words, we calculate the Berry phase accumulated by the many-body
state $\hat{\beta}_{\mathbf{k},n}^{\dagger}|S_{\mathbf{k}}\rangle$ where $|S_{\mathbf{k}}\rangle$
is the Bogoliubov vacuum. We find 
\begin{eqnarray*}
\varphi_n & = & i\oint\langle S_{\mathbf{k}}|\hat{\beta}_{\mathbf{k},n}\nabla_{\mathbf{k}}\hat{\beta}_{\mathbf{k},n}^{\dagger}|S_{\mathbf{k}}\rangle\cdot d\mathbf{k}\\
 & = & i\oint\langle S_{\mathbf{k}}|[\hat{\beta}_{\mathbf{k},n},\nabla_{\mathbf{k}}\hat{\beta}_{\mathbf{k},n}^{\dagger}]|S_{\mathbf{k}}\rangle\cdot d\mathbf{k}+i\oint\langle S_{\mathbf{k}}|\nabla_{\mathbf{k}}|S_{\mathbf{k}}\rangle\cdot d\mathbf{k}\\
 & = & i\oint\left(u_{\mathbf{k},n}^{*}[s]\nabla_{\mathbf{k}}u_{\mathbf{k},n}[s]-v_{\mathbf{k},n}^{*}[s]\nabla_{\mathbf{k}}v_{\mathbf{k},n}[s]\right)\cdot d\mathbf{k}\\
 &  & +i\oint\langle S_{\mathbf{k}}|\nabla_{\mathbf{k}}|S_{\mathbf{k}}\rangle\cdot d\mathbf{k}\\
 & = & \oint{\cal A}_n\cdot d\mathbf{k}+i\oint\langle S_{\mathbf{k}}|\nabla_{\mathbf{k}}|S_{\mathbf{k}}\rangle\cdot d\mathbf{k}.
\end{eqnarray*}
In the second line we have used that $\hat{\beta}_{\mathbf{k},n}|S_\mathbf{k}\rangle=0$
(by definition of the vacuum). We note that the Bogoliubov vacuum
$|S_\mathbf{k}\rangle$ is quasi-momentum dependend and could possibly accumulate
a Berry phase by its own, $i\oint\langle S_\mathbf{k}|\nabla_{\mathbf{k}}|S_\mathbf{k}\rangle\cdot d\mathbf{k}\neq0$.
However, the Berry phase of interest is the additional Berry phase
accumulated by the quasi-particle added over the Bogoliubov vacuum.

\subsection{Properties of the symplectic Chern numbers}

Taking into account the orthonormality condition Eq.~(\ref{eq:orthonormality}), one can immediately prove that the Chern numbers have the usual
properties: (i) They are integer numbers; (ii) After a phase transition
where two or more bands touch the individual Chern number of the band
involved in the crossings may change but their sum does not change.
Since the crossing of a particle and hole band lead to an instability
rather than a phase transition, the sum of the Chern numbers over
the particle bands is zero.

\section{Details of the calculation of the topological phase diagrams}
\label{Appendix:phasediagram}

\subsection{Symmetry of the topological phase diagram  under synthetic magnetic field inversion}
In the topological phase diagram of Fig.~\ref{fig:PhaseDiagram}(b) all Chern numbers change sign if the direction of the synthetic gauge field
is inverted, $\Phi\to-\Phi$. This has a simple explanation: To change the sign of the flux $\Phi$ and of the quasimomentum $\mathbf{k}$ corresponds to taking the complex conjugate of the BdG Hamiltonian in momentum space, $\hat{\tilde{h}}(-\mathbf{k},-\Phi)=\hat{\tilde{h}}^*(\mathbf{k},\Phi)$, c.f. Eq.~\ref{eq:BogoliubovDeGenneseff}. It follows that the single-particle eigenfunctions for opposite values of the flux and of the quasi-momentum are also related by complex conjugation, $|\mathbf{k}_n(\Phi)\rangle=(|-\mathbf{k}_n(-\Phi)\rangle)^*$. From the definition of the Chern numbers, c.~f.~Eq.~\ref{eq:Chernnumber}, it immediately follows that the Chern numbers change sign under inversion of the synthetic gauge field $\Phi$.
\subsection{Border of the different topological phases}

At a border of a topological phase transition a pair of Chern numbers
can change their values because the corresponding bands touch. Generally
speaking bands tend to repel each other rather than crossing. However,
at a lattice symmetry point this phenomenon does not necessarily occur
because the interaction of a pair of bands can be prevented by a selection
rule. In particular, at the rotational symmetry points $K$, $K'$,
and $\Gamma$, a hole with quasi-angular momentum $m$ can only be
converted into a particle with quasi-angular momentum $m+m_{\nu}$.
We note that due to inversion symmetry the bands must touch simultaneously
at the symmetry points $K$ and $K'$. We refer to the set of parameters
where the bands touch at the symmetry points $K$ and $K'$ ($\Gamma$)
as $K$-lines ($\Gamma$-lines). When also time-reversal symmetry
is present there is band crossing at all rotational symmetry points.
We refer to the set of parameters where time-reversal symmetry occurs
as ${\cal T}$-lines. In addition, a pair of bands can touch at one
of the three $M$ points where one sublattice is decoupled from the
remaining two sublattices (a particle or hole on that sublattice can
not hop on the remaining sublattices). We note that due to rotational
symmetry a pair of bands should touch simultaneously at all three
$M$ points. We refer to the set of parameters where a pair of bands touch
at the $M$ points as $M$-lines. 

In our highly symmetric system, we expect most of the crossings to
occur at a symmetry point. However, we note that accidental crossings
away from any symmetry point are not forbidden. Indeed most (but not
all) borders of the different topological phases in Fig. 2 can be
identified with ${\cal T}$-lines, $K$-lines, $\Gamma$-lines, or
$M$-lines as explained below. 

We first focus on the effective model. The vertical lines $\Phi=0,\pm3\pi$
are ${\cal T}$-lines (there is time-reversal symmetry because the
hopping amplitude $\tilde{J}$ is real). One can also easily recognize
the $M$-lines because they are horizontal. This must be the case
because the spectrum at a $M$-point where a sublattice decouples
from the remaining sublattices does not depend on the flux $\Phi$.
Indeed, there is such a horizontal line in the phase diagram of Fig.
2(b). We note that it appears for $\tilde{\nu}^{2}\approx|\tilde{J}_{ij}|\tilde{\omega}$.
Below, we show that this analytical expression holds when the $\hat{\alpha}$
quasi-particles are described by an effective particle-conserving
Hamiltonian. We can also find an analytical expression for the $K$-line
and the $\Gamma$-lines as explained below.

We initially focus on the $\Gamma$-lines.  We regard the band crossing
condition of a pair of levels with quasi-angular momentum $m$ and
$m'$, $\tilde{E}_{m}(\Gamma)=\tilde{E}_{m'}(\Gamma)$ as an implicit
equation for the parametric coupling $\tilde{\nu}$ as a function
of the flux $\Phi$. We take advantage of the analytical expression
for the spectrum at the rotational symmetry points Eq. (\ref{eq:spectrumsymeffective})
to solve this equation. For $m=0$ and $m'=-1$, we find exactly one
real positive solution 
\begin{eqnarray*}
\tilde{\nu}&=&\frac{1}{2\sqrt{3}}\times\nonumber\\
&&\left[\left(\omega_{0}-4\tilde{J}\cos\frac{\Phi}{3}\right){}^{2}-\left(\omega_{0}-4\tilde{J}\cos\frac{2\pi+\Phi}{3}\right)^{2}\right]^{1/2}
\end{eqnarray*}
in the intervals $-3\pi\leq\Phi\leq-\pi$ and $2\pi\leq\Phi\leq3\pi$.
Keeping in mind that the phase diagram is periodic with period $6\pi$,
this solution can be thought of as a single $\Gamma$-line which goes
(for increasing flux) from $\tilde{\nu}=0$ at $\Phi=2\pi$ back to
$\tilde{\nu}=0$ at $\Phi=5\pi$ ($\Phi=-\pi$). Indeed, such a line
is visible in the phase diagram of Fig. 2(b). From the implicit equations
$\tilde{E}_{m}(\mathbf{k})=\tilde{E}_{m'}(\mathbf{k})$, $\mathbf{k}=\Gamma,K$
one can find similar formulas for the remaining $\Gamma$-line and
the $K$-lines. In particular, the other $\Gamma$-line corresponds
to the crossings of the levels with angular-momentum $m=1$ and $m=0$
and goes from $\tilde{\nu}=0$ at $\Phi=\pi$ back to $\tilde{\nu}=0$
at $\Phi=4\pi$ ($\Phi=-2\pi$), see also Fig. 2(b). There is not
a third $\Gamma$-line because the levels with quasi-angular momentum
$m=1$ and $m=-1$ are degenerate only on the ${\cal T}$-lines. Likewise,
one can show that the $K$-lines go from $\tilde{\nu}=0$ at $\Phi=-\pi$
back to $\tilde{\nu}=0$ at $\Phi=\pi$ and from $\tilde{\nu}=0$
at $\Phi=-2\pi$ back to $\tilde{\nu}=0$ at $\Phi=2\pi$, respectively.
We note that the formulas for the band crossings are exact and valid
for an arbitrary ratio of $\tilde{J}/\tilde{\omega}$. However, if
$\tilde{J}/\tilde{\omega}$ is above a finite threshold the unstable
region may overlap with the band crossings and not all topological
phases will be present in the phase diagram. 

Above we have identified all lines forming the border of the different
topological phases in Fig. 2(b) except for the lines which appear
above the $M$-lines very close to the ${\cal T}$-lines and surrounds
the white areas of the topological phase diagram Fig. 2 b. These lines
correspond to accidental crossings which occur away from any symmetry
point. They enclose four different topological phases (which are not
listed in our legend for brevity). 

Next, we discuss the topological phase diagram of the original model.
For a pump circulation $m_{\nu}=1$ ($m_{\nu}=-1$), the resulting
effective flux $\Phi$ is positive (negative), see Eq.~(\ref{eq:Jeff}). Thus,  topological phase
diagram of the original model is a deformed version of the right (left) half of the effective
model phase diagram, see panel (a) of Fig. 2 for the case $m_{\nu}=1$
$(m_{\nu}=-1)$. The case $m_{\nu}=0$ on the other hand is mapped
onto the ${\cal T}$-lines of the effective diagram. 

A remarkable feature of our model is that there is only a single topological
phase for any fixed value $m_{\nu}$ of the pump circulation if the
off-diagonal parametric terms are not present ($\nu_{{\rm off}}=0)$:
$\vec{C}=(\mp1,0,\pm1)$ for $m_{\nu}=\pm1$ and $\vec{C}=(0,0,0)$
for $m_{\nu}=0$ . This is reminiscent of the anomalous Quantum Hall
effect on a Kagome lattice with nearest neighbor hoppings (OMN model)
where the topological phase is uniquely determined by the sign of
the magnetic flux piercing a triangular plaquette, $\vec{C}=(\mp1,0,\pm1)$
if the flux is positive or negative, respectively. Indeed, for small
squeezing $\nu_{{\rm on}},J\ll\omega_{0}$, the parametric interaction
effectively induces a small synthetic gauge field with a positive
flux for $m_{\nu}=1$. This can be easily seen by switching to the
effective description and neglecting the residual parametric terms.
For concreteness we consider the case $m_{\nu}=1$. For small squeezing
and $\nu_{{\rm off}}=0$ we are somewhere close to $\tilde{\nu}=0$,
$\Phi=2\pi$ inside the topological phase $\vec{C}=(-1,0,1)$ at the
bottom right corner of the effective diagram. From the above analysis
of the effective diagram we know that a topological phase transition
can occur only if we cross a $\Gamma$ or ${\cal T}$-line. However,
from the analytical solution of the spectrum at the rotational symmetry
points of the original model Eq. (\ref{eq:spectrumrotsym}) we see
that such crossings never occur on the $\nu_{{\rm off}}=0$ axis.
Thus, there in no topological phase transition even for large squeezing
if $\nu_{{\rm off}}=0$.

\subsection{Effective excitation-conserving Hamiltonian}
\label{Appendix:RWHamiltonian}

When $\tilde{\omega}$ is much larger than $|\tilde{J}|$ and $\tilde{\nu}$
one can derive an effective excitation conserving Hamiltonian. In
the regime where $\tilde{\nu}\gg|\tilde{J}|$ it is not enough to
keep the excitation conserving terms in Eq.~(\ref{eq:HNew}) but
one should also include the leading order correction in $\tilde{\nu}/\tilde{\omega}$.
We arrive at the excitation conserving Hamiltonian with next-nearest-neighbor
hoppings,
\begin{equation}
\hat{H}_{RW}=\sum_{\mathbf{j}}\bar{\omega}_{\mathbf{}}\hat{\alpha}_{\mathbf{j}}^{\dagger}\hat{\alpha}_{\mathbf{j}}-\sum_{\langle\mathbf{j,l}\rangle}\bar{J}_{\mathbf{jl}}^{({\rm 1)}}\hat{\alpha}_{\mathbf{j}}^{\dagger}\hat{\alpha}_{\mathbf{l}}-\sum_{\langle\langle\mathbf{j,l}\rangle\rangle}\bar{J}_{\mathbf{jl}}^{({\rm 2)}}\hat{\alpha}_{\mathbf{j}}^{\dagger}\hat{\alpha}_{\mathbf{l}}.\label{eq:RWAHameff}
\end{equation}
Here, $\langle\langle\mathbf{j,l}\rangle\rangle$ indicates the sum
over next-nearest-neighbor sites and 
\begin{eqnarray*}
\bar{\omega} & = & \tilde{\omega}-\frac{2\tilde{\nu}^{2}}{\tilde{\omega}},\qquad\bar{J}_{\mathbf{jl}}^{({\rm 2)}}=\frac{\tilde{\nu}^{2}}{2\tilde{\omega}_{{\rm }}}\\
\bar{J}_{\mathbf{jl}}^{({\rm 1)}} & = & \tilde{J}_{\mathbf{jl}}+\frac{\tilde{\nu}^{2}}{2\tilde{\omega}_{{\rm }}}.
\end{eqnarray*}
In this simplified picture it is straightforward to calculate the
band structure at the $M$ points and finding the bad degeneracy condition
$\tilde{\nu}^{2}=|\tilde{J}_{\mathbf{jl}}|\tilde{\omega}$ which leads
to the horizontal line in the topological phase diagram of the effective
model.

\section{Bulk-boundary correspondence}
\label{Appendix:Bulk/Boundarycorrespondence}

It is well known that in a system with a boundary, the net number
of edge states (the number of right-movers minus the number of left-movers)
in a bulk band gap is a topological invariant \cite{Hasan2010RMP}. This statement is based
on the sole assumption that the band structure and the corresponding
eigenvectors change smoothly in the presence of a local perturbation
that does not close a gap. Thus, it clearly applies to any quadratic
Hamiltonian. For the special case of an excitation-conserving insulator
or a superconductor the bulk boundary correspondence expresses such
topological invariant in terms of the Chern numbers: the net number
of edge states in a band gap coincides with the sum of the Chern numbers
of all bands below that band gap. Here, we explicitly show that the
bulk-boundary correspondence is still valid for our model where anomalous
terms are present. 

We start noticing that the wavefunctions of the RWA Hamiltonian Eq.
(\ref{eq:RWAHameff}) depend only on two parameters: the phase $\Phi/3$
of $\tilde{J}$ and the dimensionless next-nearest-neighbor coupling
$\tilde{\nu}^{2}/(|\tilde{J}|\tilde{\omega}).$ By calculating the
phase diagram as a function of these parameters [not shown] we
see that it supports all topological phases present in the topological
phase diagram of the effective Hamiltonian for the $\hat{\alpha}$-quasiparticles
{[}our full model without approximations{]}. Thus, we can continously
interpolate between the two Hamiltonians without crossing any topological
phase transition {[}by sending $\tilde{\omega}\to\infty$ while also
tuning $\Phi$ and $\tilde{\nu}^{2}/(\tilde{\omega}|\tilde{J}|)$
to stay in the same topological phase{]}. Keeping in mind that the
bulk-boundary correspondence holds for the excitation-conserving Hamiltonian
Eq. (\ref{eq:RWAHameff}) and that the net number of edge states does
not change during the interpolation [unless a gap is closed], we can conclude that such correspondence
is valid for our model even for small $\tilde{\omega}$ where the
RWA leading to Eq. (\ref{eq:RWAHameff}) is not a good approximation.

We note that the above reasoning combined with the assumption that
a continuous interpolation between any quadratic bosonic Hamiltonian
and an excitation conserving Hamiltonian is always possible without
closing any band gap, leads to the general validity of the
bulk-edge correspondence.

\section{Details of the transport calculations}
\label{Appendix:Transport}

In our transport calculations we have included photon decay. We adopt
the standard description of the dissipative dynamics of photonic systems
in terms of the Langevin equation of input-output theory \cite{Clerk2010}, for each site:
\begin{equation}
\dot{\hat{a}}_{j}=i\hbar^{-1}[\hat{H},\hat{a}_{j}]-\kappa\hat{a}_{j}/2+\sqrt{\kappa}\hat{a}_{j}^{({\rm in)}}.\label{eq:Langevin}
\end{equation}
In practice, we consider an array of detuned parametric amplifiers
with intensity decay rate $\kappa$ and add to the standard description
of each parametric amplifier the inter-cell coherent coupling described
in the main text. The last term describes the input field and includes
the vacuum fluctuations as well as the influence of an additional
probe field. The field leaking out of each cavity at site $j$ is
given by the input-output relations $\hat{a}_{j}^{(out)}=\hat{a}_{j}^{(in)}-\sqrt{\kappa}\hat{a}_{j}$.
The above formulas give an accurate description of a photonic system
where the intrinsic losses during injection and inside the system
are negligible.

 In Fig.~(3), we show the probabilities $T_{E}(\omega,l,j)$ and
$T_{I}(\omega,l,j)$ that a photon injected on site $j$ with frequency
$\omega_{{\rm in}}=\omega+\omega_{L}/2$ is transmitted 
to site $l$ elastically (at frequency $\omega+\omega_{L}/2$) or
inelastically (at frequency $\omega_{L}/2-\omega$) where it is detected.
From the Kubo formula and the input-output relations we find
\begin{eqnarray*}
T_{E}(\omega,l,j) & = & |\delta_{lj}-i\kappa\tilde{G}_{E}(\omega,l,j)|^{2}\\
T_{I}(\omega,l,j) & = & \kappa^{2}|\tilde{G}_{I}(\omega,l,j)|^{2}.
\end{eqnarray*}
They depend on the Green's function in frequency space $\tilde{G}(\omega,l,j)=\int_{-\infty}^{\infty}dte^{i\omega t}G(t,l,j)$
where 
\[
G(t,i,j)=\begin{pmatrix}G_{E}(t,l,j) & G_{I}^{*}(t,l,j)\\
G_{I}(t,l,j) & G_{E}^{*}(t,l,j)
\end{pmatrix}
\]
with the elastic and inelastic components $G_{E}(t,l,j)=-i\Theta(t)\langle[\hat{a}_{l}(t),\hat{a}_{j}^{\dagger}(0)]\rangle$
and $G_{I}(t,l,j)=-i\Theta(t)\langle[\hat{a}_{l}^{\dagger}(t),\hat{a}_{j}^{\dagger}(0)]\rangle$,
respectively. From the Kubo formula and the input-output relations
$\hat{a}_{j}^{(out)}=\hat{a}_{j}^{(in)}-\sqrt{\kappa}\hat{a}_{j}$
we find
\begin{eqnarray}
T_{E}(\omega,l,j) & = & |\delta_{lj}-i\kappa\tilde{G}_{E}(\omega,l,j)|^{2}\label{eq:TE}\\
T_{I}(\omega,l,j) & = & \kappa^{2}|\tilde{G}_{I}(\omega,l,j)|^{2}.\label{eq:TI}
\end{eqnarray}
Taking into account that 
\[
\hat{a}_{j}(t)=\sum_{n}e^{-iE[n]t}u_{n}[j]\hat{b}_{n}+e^{iE[n]t}v_{n}^{*}[j]\hat{b}_{n}^{\dagger}
\]
where $\hat{b}_{n}$ are the ladder operators of the normal modes
for a finite array of $N$ sites and $|n\rangle=(u_{n}[1],\dots,u_{n}[N],v_{n}[1],\dots,v_{n}[N])^{T}$
are the corresponding \emph{single-particle} states, the Green's function
reads 
\begin{eqnarray}
G_{E}(\omega,l,j) & = & \sum_{n}\frac{u_{n}[l]u_{n}^{*}[j]}{\omega-E[n]+i\kappa/2}-\frac{v_{n}^{*}[l]v_{n}[j]}{\omega+E[n]+i\kappa/2},\nonumber \\
\label{eq:GE}\\
G_{I}(\omega,l,j) & = & \sum_{n}\frac{v_{n}[l]u_{n}^{*}[j]}{\omega-E[n]+i\kappa/2}-\frac{u_{n}^{*}[l]v_{n}[j]}{\omega+E[n]+i\kappa/2}.\nonumber \\
\label{eq:GI}
\end{eqnarray}
We note that for a probe field inside the bandwidth of the particle
(hole) sector but far detuned from the hole (particle) sector, only
the first (second) term of the summand in Eq. (\ref{eq:GE}) and (\ref{eq:GI})
is resonant. Thus, as expected, the inelastic scattering is comparatively
larger when the probe field is in the hole band gap.

It is easy to estimate quantitatively the relative intensities of
elastically and inelastically transmitted light when the parametric
interaction of the $\hat{\alpha}$ quasiparticles is small. In this
case, it is straighforward to show that $|v_{n}[j]/u_{n}[j]|\approx\tanh r$
independent of the band $n$ and the site $j$. By putting together
Eqs.~(\ref{eq:TE},\ref{eq:TI},\ref{eq:GE},\ref{eq:GI}) and neglecting
the off-resonant terms we find that for $\omega,\tilde{\omega}\gg|\tilde{J}|,\kappa,|\omega-\tilde{\omega}|$,
\begin{eqnarray*}
T_{I}(\omega,l,j) & \approx & (\tanh r)^{2}T_{E}(\omega,l,j)\\
\approx T_{I}(-\omega,l,j) & \approx & (\coth r)^{2}T_{E}(-\omega,l,j).
\end{eqnarray*}
These analytical formulas agree quantitatively with the numerical
results shown in Fig. 4(b) {[}note that in Fig. 4(b) the transmission
at the output sites is rescaled by the overall transmission, $\sum_{l\neq j}T_{I}(\omega,l,j)+T_{E}(\omega,l,j)${]}.

Note that we have assumed, for simplicity, that there is no intrinsic absorption present. If there is intrinsic photon absorption, that will add another decay channel to the equation for the light field, but the resulting picture for the light field propagation remains unchanged except for the expected reduction in propagation length along the edge state. 


\bibliographystyle{apsrev}
\bibliography{TopologicalBib}

\begin{thebibliography}{44}
\expandafter\ifx\csname natexlab\endcsname\relax\def\natexlab#1{#1}\fi
\expandafter\ifx\csname bibnamefont\endcsname\relax
  \def\bibnamefont#1{#1}\fi
\expandafter\ifx\csname bibfnamefont\endcsname\relax
  \def\bibfnamefont#1{#1}\fi
\expandafter\ifx\csname citenamefont\endcsname\relax
  \def\citenamefont#1{#1}\fi
\expandafter\ifx\csname url\endcsname\relax
  \def\url#1{\texttt{#1}}\fi
\expandafter\ifx\csname urlprefix\endcsname\relax\def\urlprefix{URL }\fi
\providecommand{\bibinfo}[2]{#2}
\providecommand{\eprint}[2][]{\url{#2}}

\bibitem[{\citenamefont{Goldman et~al.}(2014)\citenamefont{Goldman,
  Juzeli$\bar{{\rm u}}$nas, \"{O}hberg, and Spielman}}]{Goldman2014}
\bibinfo{author}{\bibfnamefont{N.}~\bibnamefont{Goldman}},
  \bibinfo{author}{\bibfnamefont{G.}~\bibnamefont{Juzeli$\bar{{\rm u}}$nas}},
  \bibinfo{author}{\bibfnamefont{P.}~\bibnamefont{\"{O}hberg}},
  \bibnamefont{and} \bibinfo{author}{\bibfnamefont{I.~B.}
  \bibnamefont{Spielman}}, \bibinfo{journal}{Reports on Progress in Physics}
  \textbf{\bibinfo{volume}{77}}, \bibinfo{pages}{126401}
  (\bibinfo{year}{2014}).

\bibitem[{\citenamefont{Lu et~al.}(2014)\citenamefont{Lu, Joannopoulos, and
  Soljacic}}]{Lu2014}
\bibinfo{author}{\bibfnamefont{L.}~\bibnamefont{Lu}},
  \bibinfo{author}{\bibfnamefont{J.~D.} \bibnamefont{Joannopoulos}},
  \bibnamefont{and} \bibinfo{author}{\bibfnamefont{M.}~\bibnamefont{Soljacic}},
  \bibinfo{journal}{Nat Photon} \textbf{\bibinfo{volume}{8}},
  \bibinfo{pages}{821} (\bibinfo{year}{2014}), ISSN \bibinfo{issn}{1749-4885}.

\bibitem[{\citenamefont{Prodan and Prodan}(2009)}]{Prodan2009}
\bibinfo{author}{\bibfnamefont{E.}~\bibnamefont{Prodan}} \bibnamefont{and}
  \bibinfo{author}{\bibfnamefont{C.}~\bibnamefont{Prodan}},
  \bibinfo{journal}{Phys. Rev. Lett.} \textbf{\bibinfo{volume}{103}},
  \bibinfo{pages}{248101} (\bibinfo{year}{2009}).

\bibitem[{\citenamefont{Kane and Lubensky}(2013)}]{Kane2013}
\bibinfo{author}{\bibfnamefont{C.~L.} \bibnamefont{Kane}} \bibnamefont{and}
  \bibinfo{author}{\bibfnamefont{T.~C.} \bibnamefont{Lubensky}},
  \bibinfo{journal}{Nat. Phys.} \textbf{\bibinfo{volume}{10}},
  \bibinfo{pages}{39} (\bibinfo{year}{2013}), \eprint{1308.0554}.

\bibitem[{\citenamefont{Peano et~al.}(2015)\citenamefont{Peano, Brendel,
  Schmidt, and Marquardt}}]{Peano2015}
\bibinfo{author}{\bibfnamefont{V.}~\bibnamefont{Peano}},
  \bibinfo{author}{\bibfnamefont{C.}~\bibnamefont{Brendel}},
  \bibinfo{author}{\bibfnamefont{M.}~\bibnamefont{Schmidt}}, \bibnamefont{and}
  \bibinfo{author}{\bibfnamefont{F.}~\bibnamefont{Marquardt}},
  \bibinfo{journal}{Phys. Rev. X} \textbf{\bibinfo{volume}{5}},
  \bibinfo{pages}{031011} (\bibinfo{year}{2015}).

\bibitem[{\citenamefont{Yang et~al.}(2015)\citenamefont{Yang, Gao, Shi, Lin,
  Gao, Chong, and Zhang}}]{Yang2015}
\bibinfo{author}{\bibfnamefont{Z.}~\bibnamefont{Yang}},
  \bibinfo{author}{\bibfnamefont{F.}~\bibnamefont{Gao}},
  \bibinfo{author}{\bibfnamefont{X.}~\bibnamefont{Shi}},
  \bibinfo{author}{\bibfnamefont{X.}~\bibnamefont{Lin}},
  \bibinfo{author}{\bibfnamefont{Z.}~\bibnamefont{Gao}},
  \bibinfo{author}{\bibfnamefont{Y.}~\bibnamefont{Chong}}, \bibnamefont{and}
  \bibinfo{author}{\bibfnamefont{B.}~\bibnamefont{Zhang}},
  \bibinfo{journal}{Phys. Rev. Lett.} \textbf{\bibinfo{volume}{114}},
  \bibinfo{pages}{114301} (\bibinfo{year}{2015}).

\bibitem[{\citenamefont{S\"{u}sstrunk and Huber}(2015)}]{Susstrunk2015}
\bibinfo{author}{\bibfnamefont{R.}~\bibnamefont{S\"{u}sstrunk}}
  \bibnamefont{and} \bibinfo{author}{\bibfnamefont{S.~D.} \bibnamefont{Huber}},
  \bibinfo{journal}{Science} \textbf{\bibinfo{volume}{349}},
  \bibinfo{pages}{47} (\bibinfo{year}{2015}).

\bibitem[{\citenamefont{Paulose et~al.}(2015)\citenamefont{Paulose, Chen, and
  Vitelli}}]{Paulose2015}
\bibinfo{author}{\bibfnamefont{J.}~\bibnamefont{Paulose}},
  \bibinfo{author}{\bibfnamefont{B.~G.-g.} \bibnamefont{Chen}},
  \bibnamefont{and} \bibinfo{author}{\bibfnamefont{V.}~\bibnamefont{Vitelli}},
  \bibinfo{journal}{Nat. Phys.} \textbf{\bibinfo{volume}{11}},
  \bibinfo{pages}{153} (\bibinfo{year}{2015}).

\bibitem[{\citenamefont{Nash et~al.}(2015)\citenamefont{Nash, Kleckner, Read,
  Vitelli, Turner, and Irvine}}]{Nash2015}
\bibinfo{author}{\bibfnamefont{L.~M.} \bibnamefont{Nash}},
  \bibinfo{author}{\bibfnamefont{D.}~\bibnamefont{Kleckner}},
  \bibinfo{author}{\bibfnamefont{A.}~\bibnamefont{Read}},
  \bibinfo{author}{\bibfnamefont{V.}~\bibnamefont{Vitelli}},
  \bibinfo{author}{\bibfnamefont{A.~M.} \bibnamefont{Turner}},
  \bibnamefont{and} \bibinfo{author}{\bibfnamefont{W.~T.~M.}
  \bibnamefont{Irvine}}, \bibinfo{journal}{arXiv:1504.03362v2}
  (\bibinfo{year}{2015}).

\bibitem[{\citenamefont{Haldane and Raghu}(2008)}]{Raghu2008a}
\bibinfo{author}{\bibfnamefont{F.~D.~M.} \bibnamefont{Haldane}}
  \bibnamefont{and} \bibinfo{author}{\bibfnamefont{S.}~\bibnamefont{Raghu}},
  \bibinfo{journal}{Phys. Rev. Lett.} \textbf{\bibinfo{volume}{100}},
  \bibinfo{pages}{013904} (\bibinfo{year}{2008}).

\bibitem[{\citenamefont{Raghu and Haldane}(2008)}]{Raghu2008b}
\bibinfo{author}{\bibfnamefont{S.}~\bibnamefont{Raghu}} \bibnamefont{and}
  \bibinfo{author}{\bibfnamefont{F.~D.~M.} \bibnamefont{Haldane}},
  \bibinfo{journal}{Phys. Rev. A} \textbf{\bibinfo{volume}{78}},
  \bibinfo{pages}{033834} (\bibinfo{year}{2008}).

\bibitem[{\citenamefont{Koch et~al.}(2010)\citenamefont{Koch, Houck, Hur, and
  Girvin}}]{Koch2010}
\bibinfo{author}{\bibfnamefont{J.}~\bibnamefont{Koch}},
  \bibinfo{author}{\bibfnamefont{A.~A.} \bibnamefont{Houck}},
  \bibinfo{author}{\bibfnamefont{K.~L.} \bibnamefont{Hur}}, \bibnamefont{and}
  \bibinfo{author}{\bibfnamefont{S.~M.} \bibnamefont{Girvin}},
  \bibinfo{journal}{Phys. Rev. A} \textbf{\bibinfo{volume}{82}},
  \bibinfo{pages}{043811} (\bibinfo{year}{2010}).

\bibitem[{\citenamefont{Umucal\ifmmode\imath\else\i\fi{}lar and
  Carusotto}(2011)}]{Umucallar2011}
\bibinfo{author}{\bibfnamefont{R.~O.}
  \bibnamefont{Umucal\ifmmode\imath\else\i\fi{}lar}} \bibnamefont{and}
  \bibinfo{author}{\bibfnamefont{I.}~\bibnamefont{Carusotto}},
  \bibinfo{journal}{Phys. Rev. A} \textbf{\bibinfo{volume}{84}},
  \bibinfo{pages}{043804} (\bibinfo{year}{2011}).

\bibitem[{\citenamefont{Fang et~al.}(2012)\citenamefont{Fang, Yu, and
  Fan}}]{Fang2012}
\bibinfo{author}{\bibfnamefont{K.}~\bibnamefont{Fang}},
  \bibinfo{author}{\bibfnamefont{Z.}~\bibnamefont{Yu}}, \bibnamefont{and}
  \bibinfo{author}{\bibfnamefont{S.}~\bibnamefont{Fan}},
  \bibinfo{journal}{Nature Photonics} \textbf{\bibinfo{volume}{6}},
  \bibinfo{pages}{782} (\bibinfo{year}{2012}).

\bibitem[{\citenamefont{Petrescu et~al.}(2012)\citenamefont{Petrescu, Houck,
  and Le~Hur}}]{Petrescu2012}
\bibinfo{author}{\bibfnamefont{A.}~\bibnamefont{Petrescu}},
  \bibinfo{author}{\bibfnamefont{A.~A.} \bibnamefont{Houck}}, \bibnamefont{and}
  \bibinfo{author}{\bibfnamefont{K.}~\bibnamefont{Le~Hur}},
  \bibinfo{journal}{Phys. Rev. A} \textbf{\bibinfo{volume}{86}},
  \bibinfo{pages}{053804} (\bibinfo{year}{2012}).

\bibitem[{\citenamefont{Schmidt et~al.}(2015)\citenamefont{Schmidt, Kessler,
  Peano, Painter, and Marquardt}}]{Schmidt2015}
\bibinfo{author}{\bibfnamefont{M.}~\bibnamefont{Schmidt}},
  \bibinfo{author}{\bibfnamefont{S.}~\bibnamefont{Kessler}},
  \bibinfo{author}{\bibfnamefont{V.}~\bibnamefont{Peano}},
  \bibinfo{author}{\bibfnamefont{O.}~\bibnamefont{Painter}}, \bibnamefont{and}
  \bibinfo{author}{\bibfnamefont{F.}~\bibnamefont{Marquardt}},
  \bibinfo{journal}{Optica} \textbf{\bibinfo{volume}{2}}, \bibinfo{pages}{635}
  (\bibinfo{year}{2015}).

\bibitem[{\citenamefont{Hafezi et~al.}(2011)\citenamefont{Hafezi, Demler,
  Lukin, and Taylor}}]{Hafezi2011}
\bibinfo{author}{\bibfnamefont{M.}~\bibnamefont{Hafezi}},
  \bibinfo{author}{\bibfnamefont{E.~A.} \bibnamefont{Demler}},
  \bibinfo{author}{\bibfnamefont{M.~D.} \bibnamefont{Lukin}}, \bibnamefont{and}
  \bibinfo{author}{\bibfnamefont{J.~M.} \bibnamefont{Taylor}},
  \bibinfo{journal}{Nat. Phys.} \textbf{\bibinfo{volume}{7}},
  \bibinfo{pages}{907} (\bibinfo{year}{2011}).

\bibitem[{\citenamefont{Khanikaev et~al.}(2012)\citenamefont{Khanikaev,
  Mousavi, Tse, Kargarian, MacDonald, and Shvets}}]{Khanikaevphotonic2012}
\bibinfo{author}{\bibfnamefont{A.~B.} \bibnamefont{Khanikaev}},
  \bibinfo{author}{\bibfnamefont{S.~H.} \bibnamefont{Mousavi}},
  \bibinfo{author}{\bibfnamefont{W.-K.} \bibnamefont{Tse}},
  \bibinfo{author}{\bibfnamefont{M.}~\bibnamefont{Kargarian}},
  \bibinfo{author}{\bibfnamefont{A.~H.} \bibnamefont{MacDonald}},
  \bibnamefont{and} \bibinfo{author}{\bibfnamefont{G.}~\bibnamefont{Shvets}},
  \bibinfo{journal}{Nature Materials} \textbf{\bibinfo{volume}{12}},
  \bibinfo{pages}{233} (\bibinfo{year}{2012}).

\bibitem[{\citenamefont{Hafezi et~al.}(2013)\citenamefont{Hafezi, Mittal, Fan,
  Migdall, and Taylor}}]{Hafezi2013}
\bibinfo{author}{\bibfnamefont{M.}~\bibnamefont{Hafezi}},
  \bibinfo{author}{\bibfnamefont{S.}~\bibnamefont{Mittal}},
  \bibinfo{author}{\bibfnamefont{J.}~\bibnamefont{Fan}},
  \bibinfo{author}{\bibfnamefont{A.}~\bibnamefont{Migdall}}, \bibnamefont{and}
  \bibinfo{author}{\bibfnamefont{J.~M.} \bibnamefont{Taylor}},
  \bibinfo{journal}{Nature Photonics} \textbf{\bibinfo{volume}{7}},
  \bibinfo{pages}{1001} (\bibinfo{year}{2013}).

\bibitem[{\citenamefont{Kitagawa et~al.}(2012)\citenamefont{Kitagawa, Broome,
  Fedrizzi, Rudner, Berg, Kassal, Aspuru-Guzik, Demler, and
  White}}]{Kitagawa2012}
\bibinfo{author}{\bibfnamefont{T.}~\bibnamefont{Kitagawa}},
  \bibinfo{author}{\bibfnamefont{M.~A.} \bibnamefont{Broome}},
  \bibinfo{author}{\bibfnamefont{A.}~\bibnamefont{Fedrizzi}},
  \bibinfo{author}{\bibfnamefont{M.~S.} \bibnamefont{Rudner}},
  \bibinfo{author}{\bibfnamefont{E.}~\bibnamefont{Berg}},
  \bibinfo{author}{\bibfnamefont{I.}~\bibnamefont{Kassal}},
  \bibinfo{author}{\bibfnamefont{A.}~\bibnamefont{Aspuru-Guzik}},
  \bibinfo{author}{\bibfnamefont{E.}~\bibnamefont{Demler}}, \bibnamefont{and}
  \bibinfo{author}{\bibfnamefont{A.~G.} \bibnamefont{White}},
  \bibinfo{journal}{Nat Commun} \textbf{\bibinfo{volume}{3}},
  \bibinfo{pages}{882} (\bibinfo{year}{2012}).

\bibitem[{\citenamefont{Rechtsman
  et~al.}(2013{\natexlab{a}})\citenamefont{Rechtsman, Plotnik, Zeuner, Song,
  Chen, Szameit, and Segev}}]{Rechtsman2013}
\bibinfo{author}{\bibfnamefont{M.~C.} \bibnamefont{Rechtsman}},
  \bibinfo{author}{\bibfnamefont{Y.}~\bibnamefont{Plotnik}},
  \bibinfo{author}{\bibfnamefont{J.~M.} \bibnamefont{Zeuner}},
  \bibinfo{author}{\bibfnamefont{D.}~\bibnamefont{Song}},
  \bibinfo{author}{\bibfnamefont{Z.}~\bibnamefont{Chen}},
  \bibinfo{author}{\bibfnamefont{A.}~\bibnamefont{Szameit}}, \bibnamefont{and}
  \bibinfo{author}{\bibfnamefont{M.}~\bibnamefont{Segev}},
  \bibinfo{journal}{Phys. Rev. Lett.} \textbf{\bibinfo{volume}{111}},
  \bibinfo{pages}{103901} (\bibinfo{year}{2013}{\natexlab{a}}).

\bibitem[{\citenamefont{Bardyn and \ifmmode \dot{I}\else
  \.{I}\fi{}mamo\ifmmode~\check{g}\else \v{g}\fi{}lu}(2012)}]{Bardyn2012}
\bibinfo{author}{\bibfnamefont{C.-E.} \bibnamefont{Bardyn}} \bibnamefont{and}
  \bibinfo{author}{\bibfnamefont{A.}~\bibnamefont{\ifmmode \dot{I}\else
  \.{I}\fi{}mamo\ifmmode~\check{g}\else \v{g}\fi{}lu}}, \bibinfo{journal}{Phys.
  Rev. Lett.} \textbf{\bibinfo{volume}{109}}, \bibinfo{pages}{253606}
  (\bibinfo{year}{2012}).

\bibitem[{\citenamefont{Wang et~al.}(2009)\citenamefont{Wang, Chong,
  Joannopoulos, and Soljacic}}]{Wang2009}
\bibinfo{author}{\bibfnamefont{Z.}~\bibnamefont{Wang}},
  \bibinfo{author}{\bibfnamefont{Y.}~\bibnamefont{Chong}},
  \bibinfo{author}{\bibfnamefont{J.~D.} \bibnamefont{Joannopoulos}},
  \bibnamefont{and} \bibinfo{author}{\bibfnamefont{M.}~\bibnamefont{Soljacic}},
  \bibinfo{journal}{Nature} \textbf{\bibinfo{volume}{461}},
  \bibinfo{pages}{772} (\bibinfo{year}{2009}).

\bibitem[{\citenamefont{Rechtsman
  et~al.}(2013{\natexlab{b}})\citenamefont{Rechtsman, Zeuner, Plotnik, Lumer,
  Podolsky, Dreisow, Nolte, Segev, and Szameit}}]{Rechtsman2013b}
\bibinfo{author}{\bibfnamefont{M.~C.} \bibnamefont{Rechtsman}},
  \bibinfo{author}{\bibfnamefont{J.~M.} \bibnamefont{Zeuner}},
  \bibinfo{author}{\bibfnamefont{Y.}~\bibnamefont{Plotnik}},
  \bibinfo{author}{\bibfnamefont{Y.}~\bibnamefont{Lumer}},
  \bibinfo{author}{\bibfnamefont{D.}~\bibnamefont{Podolsky}},
  \bibinfo{author}{\bibfnamefont{F.}~\bibnamefont{Dreisow}},
  \bibinfo{author}{\bibfnamefont{S.}~\bibnamefont{Nolte}},
  \bibinfo{author}{\bibfnamefont{M.}~\bibnamefont{Segev}}, \bibnamefont{and}
  \bibinfo{author}{\bibfnamefont{A.}~\bibnamefont{Szameit}},
  \bibinfo{journal}{Nature} \textbf{\bibinfo{volume}{496}},
  \bibinfo{pages}{196} (\bibinfo{year}{2013}{\natexlab{b}}).

\bibitem[{\citenamefont{Mittal et~al.}(2014)\citenamefont{Mittal, Fan, Faez,
  Migdall, Taylor, and Hafezi}}]{Mittal2014}
\bibinfo{author}{\bibfnamefont{S.}~\bibnamefont{Mittal}},
  \bibinfo{author}{\bibfnamefont{J.}~\bibnamefont{Fan}},
  \bibinfo{author}{\bibfnamefont{S.}~\bibnamefont{Faez}},
  \bibinfo{author}{\bibfnamefont{A.}~\bibnamefont{Migdall}},
  \bibinfo{author}{\bibfnamefont{J.~M.} \bibnamefont{Taylor}},
  \bibnamefont{and} \bibinfo{author}{\bibfnamefont{M.}~\bibnamefont{Hafezi}},
  \bibinfo{journal}{Phys. Rev. Lett.} \textbf{\bibinfo{volume}{113}},
  \bibinfo{pages}{087403} (\bibinfo{year}{2014}).

\bibitem[{\citenamefont{Tzuang et~al.}(2014)\citenamefont{Tzuang, Fang,
  Nussenzveig, Fan, and Lipson}}]{2014_Lipson_NonreciprocalPhaseShift}
\bibinfo{author}{\bibfnamefont{L.~D.} \bibnamefont{Tzuang}},
  \bibinfo{author}{\bibfnamefont{K.}~\bibnamefont{Fang}},
  \bibinfo{author}{\bibfnamefont{P.}~\bibnamefont{Nussenzveig}},
  \bibinfo{author}{\bibfnamefont{S.}~\bibnamefont{Fan}}, \bibnamefont{and}
  \bibinfo{author}{\bibfnamefont{M.}~\bibnamefont{Lipson}},
  \bibinfo{journal}{Nature Photonics} \textbf{\bibinfo{volume}{8}},
  \bibinfo{pages}{701} (\bibinfo{year}{2014}).

\bibitem[{\citenamefont{Shindou et~al.}(2013)\citenamefont{Shindou, Matsumoto,
  Murakami, and Ohe}}]{Shindou2013}
\bibinfo{author}{\bibfnamefont{R.}~\bibnamefont{Shindou}},
  \bibinfo{author}{\bibfnamefont{R.}~\bibnamefont{Matsumoto}},
  \bibinfo{author}{\bibfnamefont{S.}~\bibnamefont{Murakami}}, \bibnamefont{and}
  \bibinfo{author}{\bibfnamefont{J.-i.} \bibnamefont{Ohe}},
  \bibinfo{journal}{Phys. Rev. B} \textbf{\bibinfo{volume}{87}},
  \bibinfo{pages}{174427} (\bibinfo{year}{2013}).

\bibitem[{\citenamefont{Engelhardt and Brandes}(2015)}]{Brandes2015}
\bibinfo{author}{\bibfnamefont{G.}~\bibnamefont{Engelhardt}} \bibnamefont{and}
  \bibinfo{author}{\bibfnamefont{T.}~\bibnamefont{Brandes}},
  \bibinfo{journal}{arXiv:1503.02503v1}  (\bibinfo{year}{2015}).

\bibitem[{\citenamefont{Bardyn et~al.}(2015)\citenamefont{Bardyn, Karzig,
  Refael, and Liew}}]{Liew2015}
\bibinfo{author}{\bibfnamefont{C.-E.} \bibnamefont{Bardyn}},
  \bibinfo{author}{\bibfnamefont{T.}~\bibnamefont{Karzig}},
  \bibinfo{author}{\bibfnamefont{G.}~\bibnamefont{Refael}}, \bibnamefont{and}
  \bibinfo{author}{\bibfnamefont{T.~C.~H.} \bibnamefont{Liew}},
  \bibinfo{journal}{arXiv:1503.08824v1}  (\bibinfo{year}{2015}).

\bibitem[{\citenamefont{Ohgushi et~al.}(2000)\citenamefont{Ohgushi, Murakami,
  and Nagaosa}}]{Ohgushi2000}
\bibinfo{author}{\bibfnamefont{K.}~\bibnamefont{Ohgushi}},
  \bibinfo{author}{\bibfnamefont{S.}~\bibnamefont{Murakami}}, \bibnamefont{and}
  \bibinfo{author}{\bibfnamefont{N.}~\bibnamefont{Nagaosa}},
  \bibinfo{journal}{Phys. Rev. B} \textbf{\bibinfo{volume}{62}},
  \bibinfo{pages}{R6065} (\bibinfo{year}{2000}).

\bibitem[{\citenamefont{Green et~al.}(2010)\citenamefont{Green, Santos, and
  Chamon}}]{Green2010}
\bibinfo{author}{\bibfnamefont{D.}~\bibnamefont{Green}},
  \bibinfo{author}{\bibfnamefont{L.}~\bibnamefont{Santos}}, \bibnamefont{and}
  \bibinfo{author}{\bibfnamefont{C.}~\bibnamefont{Chamon}},
  \bibinfo{journal}{Phys. Rev. B} \textbf{\bibinfo{volume}{82}},
  \bibinfo{pages}{075104} (\bibinfo{year}{2010}).

\bibitem[{\citenamefont{Fukui et~al.}(2005)\citenamefont{Fukui, Hatsugai, and
  Suzuki}}]{Fukui2005}
\bibinfo{author}{\bibfnamefont{T.}~\bibnamefont{Fukui}},
  \bibinfo{author}{\bibfnamefont{Y.}~\bibnamefont{Hatsugai}}, \bibnamefont{and}
  \bibinfo{author}{\bibfnamefont{H.}~\bibnamefont{Suzuki}},
  \bibinfo{journal}{Journal of the Physical Society of Japan}
  \textbf{\bibinfo{volume}{74}}, \bibinfo{pages}{1674} (\bibinfo{year}{2005}).

\bibitem[{\citenamefont{Mookherjea and Yariv}(2002)}]{Yariv2002}
\bibinfo{author}{\bibfnamefont{S.}~\bibnamefont{Mookherjea}} \bibnamefont{and}
  \bibinfo{author}{\bibfnamefont{A.}~\bibnamefont{Yariv}},
  \bibinfo{journal}{Ieee J Quantum Elect} \textbf{\bibinfo{volume}{8}},
  \bibinfo{pages}{448} (\bibinfo{year}{2002}).

\bibitem[{\citenamefont{Eggleton et~al.}(2011)\citenamefont{Eggleton,
  Luther-Davies, and Richardson}}]{Eggleton2011}
\bibinfo{author}{\bibfnamefont{B.~J.} \bibnamefont{Eggleton}},
  \bibinfo{author}{\bibfnamefont{B.}~\bibnamefont{Luther-Davies}},
  \bibnamefont{and}
  \bibinfo{author}{\bibfnamefont{K.}~\bibnamefont{Richardson}},
  \bibinfo{journal}{Nat Photonics} \textbf{\bibinfo{volume}{5}},
  \bibinfo{pages}{141} (\bibinfo{year}{2011}).

\bibitem[{\citenamefont{Dahdah et~al.}(2011)\citenamefont{Dahdah, Pilar-Bernal,
  Courjal, Ulliac, and Baida}}]{Dahdah2011}
\bibinfo{author}{\bibfnamefont{J.}~\bibnamefont{Dahdah}},
  \bibinfo{author}{\bibfnamefont{M.}~\bibnamefont{Pilar-Bernal}},
  \bibinfo{author}{\bibfnamefont{N.}~\bibnamefont{Courjal}},
  \bibinfo{author}{\bibfnamefont{G.}~\bibnamefont{Ulliac}}, \bibnamefont{and}
  \bibinfo{author}{\bibfnamefont{F.}~\bibnamefont{Baida}}, \bibinfo{journal}{J.
  Appl. Phys.} \textbf{\bibinfo{volume}{110}},  (\bibinfo{year}{2011}).

\bibitem[{\citenamefont{Safavi-Naeini et~al.}(2013)\citenamefont{Safavi-Naeini,
  Gr{\"o}blacher, Hill, Chan, Aspelmeyer, and Painter}}]{Painter2013}
\bibinfo{author}{\bibfnamefont{A.~H.} \bibnamefont{Safavi-Naeini}},
  \bibinfo{author}{\bibfnamefont{S.}~\bibnamefont{Gr{\"o}blacher}},
  \bibinfo{author}{\bibfnamefont{J.~T.} \bibnamefont{Hill}},
  \bibinfo{author}{\bibfnamefont{J.}~\bibnamefont{Chan}},
  \bibinfo{author}{\bibfnamefont{M.}~\bibnamefont{Aspelmeyer}},
  \bibnamefont{and} \bibinfo{author}{\bibfnamefont{O.}~\bibnamefont{Painter}},
  \bibinfo{journal}{Nature} \textbf{\bibinfo{volume}{500}},
  \bibinfo{pages}{185} (\bibinfo{year}{2013}).

\bibitem[{\citenamefont{Purdy et~al.}(2013)\citenamefont{Purdy, Yu, Peterson,
  Kampel, and Regal}}]{Regal2013}
\bibinfo{author}{\bibfnamefont{T.~P.} \bibnamefont{Purdy}},
  \bibinfo{author}{\bibfnamefont{P.~L.} \bibnamefont{Yu}},
  \bibinfo{author}{\bibfnamefont{R.~W.} \bibnamefont{Peterson}},
  \bibinfo{author}{\bibfnamefont{N.~S.} \bibnamefont{Kampel}},
  \bibnamefont{and} \bibinfo{author}{\bibfnamefont{C.~A.} \bibnamefont{Regal}},
  \bibinfo{journal}{Phys. Rev. X} \textbf{\bibinfo{volume}{3}},
  \bibinfo{pages}{031012} (\bibinfo{year}{2013}).

\bibitem[{\citenamefont{Kronwald et~al.}(2013)\citenamefont{Kronwald,
  Marquardt, and Clerk}}]{Kronwald2013}
\bibinfo{author}{\bibfnamefont{A.}~\bibnamefont{Kronwald}},
  \bibinfo{author}{\bibfnamefont{F.}~\bibnamefont{Marquardt}},
  \bibnamefont{and} \bibinfo{author}{\bibfnamefont{A.~A.} \bibnamefont{Clerk}},
  \bibinfo{journal}{Phys. Rev. A} \textbf{\bibinfo{volume}{88}},
  \bibinfo{pages}{063833} (\bibinfo{year}{2013}).

\bibitem[{\citenamefont{Bergeal et~al.}(2010)\citenamefont{Bergeal, Vijay,
  Manucharyan, Siddiqi, Schoelkopf, Girvin, and Devoret}}]{Bergeal2010}
\bibinfo{author}{\bibfnamefont{N.}~\bibnamefont{Bergeal}},
  \bibinfo{author}{\bibfnamefont{R.}~\bibnamefont{Vijay}},
  \bibinfo{author}{\bibfnamefont{V.~E.} \bibnamefont{Manucharyan}},
  \bibinfo{author}{\bibfnamefont{I.}~\bibnamefont{Siddiqi}},
  \bibinfo{author}{\bibfnamefont{R.~J.} \bibnamefont{Schoelkopf}},
  \bibinfo{author}{\bibfnamefont{S.~M.} \bibnamefont{Girvin}},
  \bibnamefont{and} \bibinfo{author}{\bibfnamefont{M.}~\bibnamefont{Devoret}},
  \bibinfo{journal}{Nat. Phys.} \textbf{\bibinfo{volume}{6}},
  \bibinfo{pages}{296} (\bibinfo{year}{2010}).

\bibitem[{\citenamefont{Abdo et~al.}(2013)\citenamefont{Abdo, Kamal, and
  Devoret}}]{Abdo2013}
\bibinfo{author}{\bibfnamefont{B.}~\bibnamefont{Abdo}},
  \bibinfo{author}{\bibfnamefont{A.}~\bibnamefont{Kamal}}, \bibnamefont{and}
  \bibinfo{author}{\bibfnamefont{M.}~\bibnamefont{Devoret}},
  \bibinfo{journal}{Phys. Rev. B} \textbf{\bibinfo{volume}{87}},
  \bibinfo{pages}{014508} (\bibinfo{year}{2013}).

\bibitem[{\citenamefont{Gurarie and Chalker}(2003)}]{Gurarie2003}
\bibinfo{author}{\bibfnamefont{V.}~\bibnamefont{Gurarie}} \bibnamefont{and}
  \bibinfo{author}{\bibfnamefont{J.~T.} \bibnamefont{Chalker}},
  \bibinfo{journal}{Phys. Rev. B} \textbf{\bibinfo{volume}{68}},
  \bibinfo{pages}{134207} (\bibinfo{year}{2003}).

\bibitem[{\citenamefont{Ryu et~al.}(2010)\citenamefont{Ryu, Schnyder, Furusaki,
  and Ludwig}}]{Ryu2010}
\bibinfo{author}{\bibfnamefont{S.}~\bibnamefont{Ryu}},
  \bibinfo{author}{\bibfnamefont{A.~P.} \bibnamefont{Schnyder}},
  \bibinfo{author}{\bibfnamefont{A.}~\bibnamefont{Furusaki}}, \bibnamefont{and}
  \bibinfo{author}{\bibfnamefont{A.~W.~W.} \bibnamefont{Ludwig}},
  \bibinfo{journal}{New J. Phys.} \textbf{\bibinfo{volume}{12}},
  \bibinfo{pages}{065010} (\bibinfo{year}{2010}).

\bibitem[{\citenamefont{Hasan and Kane}(2010)}]{Hasan2010RMP}
\bibinfo{author}{\bibfnamefont{M.~Z.} \bibnamefont{Hasan}} \bibnamefont{and}
  \bibinfo{author}{\bibfnamefont{C.~L.} \bibnamefont{Kane}},
  \bibinfo{journal}{Rev. Mod. Phys.} \textbf{\bibinfo{volume}{82}},
  \bibinfo{pages}{3045} (\bibinfo{year}{2010}).

\bibitem[{\citenamefont{Clerk et~al.}(2010)\citenamefont{Clerk, Devoret,
  Girvin, Marquardt, and Schoelkopf}}]{Clerk2010}
\bibinfo{author}{\bibfnamefont{A.~A.} \bibnamefont{Clerk}},
  \bibinfo{author}{\bibfnamefont{M.~H.} \bibnamefont{Devoret}},
  \bibinfo{author}{\bibfnamefont{S.~M.} \bibnamefont{Girvin}},
  \bibinfo{author}{\bibfnamefont{F.}~\bibnamefont{Marquardt}},
  \bibnamefont{and} \bibinfo{author}{\bibfnamefont{R.~J.}
  \bibnamefont{Schoelkopf}}, \bibinfo{journal}{Reviews of Modern Physics}
  \textbf{\bibinfo{volume}{82}}, \bibinfo{pages}{1155} (\bibinfo{year}{2010}),
  \eprint{0810.4729}.

\end{thebibliography}

\end{document}